\documentclass[twocolumn,showpacs,aps,epsfig]{revtex4}

\usepackage{graphicx}
\usepackage{epstopdf}
\usepackage[center]{subfigure}

\begin{document}

%%%%%%%%%%%%%%%%%%%%%%%%%%%%%%%%%%%%%%%%%%%%%%%%%%%%%%%%%%%%%%%
 \newcommand{\bq}{\begin{equation}}
 \newcommand{\eq}{\end{equation}}
 \newcommand{\bqn}{\begin{eqnarray}}
 \newcommand{\eqn}{\end{eqnarray}}
 \newcommand{\nb}{\nonumber}
 \newcommand{\lb}{\label}
 %%%%%%%%%%%%%%%%%%%%%%%%%%%%%%%%%%%%%%%%%%%%%%%%%%%%%%%%%%%%%%%

\title{Two 3-Branes in Randall-Sundrum Setup and Current Acceleration 
of the Universe}
\author{  Anzhong Wang   
\thanks{Electronic address:  Anzhong$\_$Wang@baylor.edu}  }
\affiliation{  
CASPER, Department of Physics, Baylor University,
Waco, Texas 76798-7316}
\author{Rong-Gen Cai \thanks{E-mail address:
cairg@itp.ac.cn}}
\affiliation{
  Institute of Theoretical Physics, Chinese
Academy of Sciences,
 P.O. Box 2735, Beijing 100080, China}
\author{N.O. Santos \thanks{E-mail: nos@cbpf.br and santos@ccr.jussieu.fr}}
\affiliation{ School of Mathematical Sciences, Queen Mary,
University of London, London E1 4NS, UK\\
 LERMA/CNRS-FRE 2460, Universit\'e Pierre et Marie Curie, ERGA, 
Bo\^{\i}te 142, 4 Place Jussieu, 75005 Paris Cedex 05,
France\\
Laborat\'orio Nacional de Computa\c{c}\~{a}o Cient\'{\i}fica, 
25651-070 Petr\'opolis RJ, Brazil}

\date{\today}

\begin{abstract}

Five-dimensional spacetimes of two orbifold 3-branes are studied, 
by assuming that {\em the two 3-branes are 
spatially homogeneous, isotropic, and independent of time}, following the 
so-called ``bulk-based" approach. The most general form of the metric is 
obtained, and the corresponding field equations are divided into three 
groups, one is valid on each of the two 3-branes, and the third is valid 
in the bulk. The Einstein tensor on the 3-branes is expressed in 
terms of the discontinuities of the first-order derivatives of the
metric coefficients. Thus, once the metric is known in the bulk, the 
distribution of the Einstein tensor on the two 3-branes is uniquely 
determined. As applications, we consider two different cases, one is in 
which the bulk is locally $AdS_{5}$, and the other is where it is vacuum. 
In some cases, it is shown that the universe is first decelerating and then 
accelerating.  The global structure of the bulk as well as the 3-branes
is also studied, and found that in some cases the solutions may represent 
the collision of two orbifold 3-branes. The applications of the formulas 
to the studies of the cyclic universe  and the cosmological constant
problem are also pointed out.

\end{abstract}

\vspace{.7cm}


\maketitle

%%%%%%%%%%%%%%%%%%%%%%%%%%%%%%%%%%%%%%%%%%%%%%%%%%%%%%%%%%%%%%%%%%%%%%%%%%%%%%%%%
\vspace{1.cm}

\section{Introduction}

\renewcommand{\theequation}{1.\arabic{equation}}
\setcounter{equation}{0}

Superstring and M-theory all suggest that we may live in a world that 
has more than three spatial
dimensions.  Because only three of these are presently observable, one has to
explain why the others are hidden from detection.  One such explanation
is the so-called Kaluza-Klein (KK) compactification, according to which the
size of the extra dimensions is very small (often taken to be on the order
of the Planck length).  As a consequence, modes that have momentum in the
directions of the extra dimensions are excited at currently inaccessible
energies.

Recently, the braneworld scenarios \cite{ADD,RS1} has dramatically changed
this point of view and, in the process, received a great deal of attention.
At present, there are a number of proposed models  (See, for example,
\cite{reviews} and references therein.). In particular,
Arkani-Hamed {\em et al} (ADD) \cite{ADD} pointed out that the extra
dimensions need not necessarily be small and may even be in the scale of
millimeters \cite{Hoyle04}.  This model assumes that Standard Model fields 
are confined to a three (spatial) dimensional surface (a 3-brane) living in a
larger dimensional bulk, while the gravitational field propagates
in the whole bulk.  Additional fields may live only on the brane or
in the whole bulk, provided that their current undetectability is 
consistent with experimental bounds. One of the most attractive features 
of this model is that it may potentially resolve the long standing 
{\em hierarchy problem}, namely the large difference in magnitudes
between the Planck and electroweak scales. 

An alternative approach was proposed by Randall and Sundrum (RS) \cite{RS1},
which will be referred to as RS1 model.
One of the most attractive features of this model is that it will soon be 
explored by LHC \cite{DHR00}. For critical reviews of the model and some 
open issues, we refer readers to \cite{RC04}. The spacetime in this model is 
five-dimensional, with the extra dimension being compactified on an 
$S^{1}/Z_{2}$ orbifold, that is, the extra dimension is periodic,
$\phi \in[-\pi, \pi]$, and its points with ($x^{\mu}, \phi$) are identified 
with the ones ($x^{\mu}, -\phi$). In such a setting,  the spacetime necessarily 
contains two 3-branes, located, respectively, at the fixed points $\phi = 0$, 
and $\phi = \pi$. The   brane at $\phi = 0$ is usually called hidden (or Planck) 
brane, and the  one at $\phi = \pi$ is called visible (or TeV) brane. 
The corresponding 5D spacetime is locally anti-de Sitter ($AdS_{5}$), and  
described by the metric,
\bq
\lb{1.5}
ds^{2}_{5} = e^{-2kr_{c}|\phi|}\eta_{\mu\nu} dx^{\mu}dx^{\nu}
- r^{2}_{c}d\phi^{2},
\eq
for which the 4D spacetimes of the two 3-branes are Poincar\'e invariant, 
where $\mu,\; \nu = 0, 1, 2, 3$, and $k$ is a constant and of the order of 
$M$, where $M$ is the five-dimensional mass scale. $\eta_{\mu\nu}$ denotes 
the 4D Minkowski metric,
and $r_{c}$ the radius of the extra dimension.  To obtain the relationship
between $M$ and the four-dimensional Planck scale $M_{pl}$, let us consider 
the gravitational perturbations, given by
\bq
\lb{1.6}
ds^{2}_{5} = e^{-2kr_{c}|\phi|}\left[\eta_{\mu\nu} 
+ h_{\mu\nu}\left(x^{\lambda}\right)\right] dx^{\mu}dx^{\nu}
- r^{2}_{c}d\phi^{2},
\eq
we find that
\bqn
\lb{1.7}
S^{(5)}_{g} &=& \frac{M^{3}}{16\pi}\int{dx^{4} d\phi\sqrt{ g^{(5)}}
R^{(5)}}\nb\\
 &=&  \frac{M^{3}}{16\pi}\int{d^{4}x}\int^{\pi}_{-\pi}{
d\phi r_{c}e^{-2kr_{c}|\phi|} \sqrt{- g^{(4)}}
R^{(4)}},\nb\\
\eqn
where $g^{(4)} \equiv \eta_{\mu\nu} + h_{\mu\nu}$. Then, comparing Eq.(\ref{1.7})
with  
\bq
\label{1.1a}
S_{g}^{(4)}  = \frac{M^{2}_{pl} }{16\pi}\int{d^{4}x  \sqrt{- g^{(4)}}
R^{(4)}},
\eq
%where $M_{pl}$ is the observed four-dimensional Planck mass with $M_{pl} 
%\sim 10^{16} \; TeV$, 
we find that 
\bq
\lb{1.8}
M^{2}_{pl} = \frac{M^{3}}{k}\left(1 - e^{-2kr_{c}\pi}\right).
\eq
Thus, in the RS1 model, for large $kr_{c}$ the Planck $M_{pl}$ is weakly 
dependent on $r_{c}$ 
and $M_{pl} \sim M$. Therefore, in contrast to the ADD model, 
the RS1 model predicts that $M$ and $M_{pl}$ are in the same order,
$M \sim M_{pl} \sim 10^{16} \; TeV$. The resolution of the hierarchy problem
comes from the warped factor $e^{-2kr_{c}\pi}$ on the visible brane.
To show this explicitly, let us consider a scalar field $H$ confined
on the brane,
\bqn
\lb{1.9}
S_{eff.}^{vis} &=& \int_{\phi=\pi}
{d^{4}x  \sqrt{-g_{vis}}\left[g^{\mu\nu}_{vis}
\nabla_{\mu}H^{*}\nabla_{\nu}H  \right.}\nb\\
& & {\left. - \lambda\left(\left|H\right|^{2}
- m^{2}_{0}\right)^{2}\right]}\nb\\
& = & \int_{\phi=\pi}
{d^{4}x   \left[g^{\mu\nu}
\nabla_{\mu}{\bar{H}}^{*}\nabla_{\nu}{\bar{H}}  \right.}\nb\\
& & \left.
- \lambda\left(\left|{\bar{H}}\right|^{2}
- e^{-2k\pi r_{c}}m^{2}_{0}\right)^{2}\right],
\eqn
where ${\bar{H}} \equiv e^{k\pi r_{c}}H$, and
\bq
\lb{1.10}
g_{vis}^{\mu\nu} \equiv g_{5}^{\mu\nu}(x, \pi)
= e^{2k\pi r_{c}} g^{\mu\nu}.
\eq
Therefore, the mass measured by $g_{\mu\nu}$ is given by
\bq
\lb{1.11}
m = e^{-k\pi r_{c}}m_{0}. 
\eq
Then, for $k r_{c} \sim  11$ and $m_{0} \sim  10^{16} \; TeV$, we find  
 $m \sim  TeV$.
 
When extra dimensions exist, a critical ingredient is the stability of the
extra dimensions. Goldberger and Wise showed that values of  $k r_{c} 
\sim 11$ are indeed natural and can be provided by a stable configuration
\cite{GW99}.

In this paper, we  study cosmological models in the RS1 setup. 
In the original model,
only gravity propagated in the bulk and the 
standard model fields were confined
to the TeV brane. As a results,  the bulk is necessarily 
a generalized Schwarzschild-anti-de
Sitter space \cite{KI99}. However, soon it was realized 
that  much richer phenomena 
can be obtained, if some of the standard model fields 
are allowed to propagate in the bulk 
\cite{DHR00}. In order to incorporate all these 
situations, we shall allow the bulk to be 
filled with any matter field(s). 

In addition, much of the work in brane cosmology 
has taken the so-called ``brane-based" 
approach \cite{reviews,RS1CMs}. In this paper, 
we shall follow the   ``bulk-based" 
approach \cite{BCG00}, initiated by the authors 
of \cite{KI99}. In \cite{BCG00} spacetimes 
with one 3-brane were studied systematically by 
the ``bulk-based" approach, and in particular
the most general form of the bulk metric was found. 
In this paper, we shall generalize those 
studies to the case of two 3-branes. As shown explicitly 
in Sec. II, their general form of the 
bulk metric is a particular case of ours for two 3-branes. 

The rest of the paper is organized as follows: In Sec. II, starting with the 
assumption that {\em the two orbifold 3-branes are 
spatially homogeneous, isotropic, and independent 
of time, and that the extra dimension has $S^{1}/Z_{2}$ symmetry}, we  derive the 
most general form of the metric. It is very important to note that the orbifold
symmetry manifests itself explicitly only in  certain coordinate system. In this
paper, we take the point of view that this particular coordinate system is the one
in which the two 3-branes are all at rest. With this in
mind,  using gauge freedom we first map the two 3-branes into fixed points, in
which  the proper distance between the two 3-branes in general is time-dependent. 
Then, using distribution theory we develop a general formula for two 3-branes 
with orbifold symmetry. In particular, we divide the field equations into three 
different groups, given, respectively, by Eqs.(\ref{2.18a}), (\ref{2.18b}) and 
(\ref{2.18c}), where Eq.(\ref{2.18a}) holds in the bulk, and Eqs.(\ref{2.18b}) and 
(\ref{2.18c})  hold on each of the two 3-branes. 
The Einstein tensor on the two 3-branes is given in terms of the discontinuities of the 
metric coefficients. Thus, once the metric is known in   the bulk, 
$G^{(0)}_{AB}$ and $G^{(c)}_{AB}$ are known, and   Eqs.(\ref{2.18b}) and (\ref{2.18c}) 
will uniquely  determine the distribution of the matter fields on the two 3-branes. 
In Sec. III, as   applications of our general formulas, we consider two 3-branes in
an $AdS_{5}$  bulk, by imposing the conditions that the cosmological constant on 
the visible brane vanishes and the equation of state of its matter field
is given by $p_{c} = w \rho_{c}$, with $w$ being a constant. From this example we can see
that different ``cut-paste" operations result in different models, even although 
the bulks are all locally anti-de Sitter. The global structure of the bulk and
the 3-branes is also studied, and   found that in some cases the geodesically
complete spacetime represents infinite number of 3-branes. In Sec. IV, similar 
considerations are 
carried out, but now the bulk is vacuum. After re-deriving the general vacuum solutions
of the five-dimensional bulk \cite{BCG00}, we apply our general formulas to these
solutions for two 3-branes, and whereby obtain the conditions under which the universe 
undergoes a current acceleration.   The study of global structure of these solutions
shows that some solutions may represent the collision of two orbifold 3-branes. 
Sec. V contains our main conclusions and remarks.  

It should be noted that the problem has been studied so intensively, there is 
inevitably overlap between previous works and what we present here. However, as
far as we know, it is the first time to present such a general treatment of two
orbifold 3branes in arbitrary bulks and systematically study the global structures 
of the bulk for all the solutions with/without the bulk cosmological constant.

\section{Gauge Freedom and Gauge Choice For Two Homogeneous and Isotropic 3-Branes}

\renewcommand{\theequation}{2.\arabic{equation}}
\setcounter{equation}{0}

In this paper we consider spacetimes that are five-dimensional 
and contain two 3-branes. 
Since we shall apply such spacetimes to cosmology, we further 
assume that the two 3-branes
are spatially homogeneous, isotropic, and independent of time.  
The fifth dimension is periodic 
and has a reflection (orbifold) symmetry with respect to each 
of the two 3-branes. Then, one can see that the space between 
the two 3-branes represents half of the periodic space along the 
fifth dimension. 

To start with, let us first consider the conditions 
that {\em the space on the two 
3-branes is spatially homogeneous, isotropic, and independent 
of time}.  It is not difficult to 
show that such a space must have a constant curvature  
at any point of the two 3-branes,
and its metric   must  take the form \cite{Wolf},
\bq 
\lb{2.1}
d\Sigma^{2}_{k} \equiv \gamma_{ab} dx^{a}dx^{b} 
= \frac{dr^{2}}{1 - k r^{2}} + r^{2}d\Omega^{2},
\eq
where $ d\Omega^{2} \equiv d\theta^{2} + \sin^{2}\theta d\phi^{2}$, 
and $ a, b = 1, 2, 3$. The constant $k$  represents the curvature of the 3-space,
and can be positive, negative or zero. 
%%%%%%%%%%%%%%%%%%%%%%%%%%%%%%%%%%%%%%%%%%%%%%%%%%%%%%%%%%%%%%%%%%%%%%%%%%%%%%%%%%%%%%%%%%%%%%
%, with the properties
%\bqn
%\lb{2.2}
%R_{abcd} &=& k\left(\gamma_{ac} \gamma_{bd} - \gamma_{ac} \gamma_{bd}\right),\nb\\
%R_{bd} &\equiv& \gamma^{ac}R_{abcd} = 2k \gamma_{bd}.
%\eqn
%%%%%%%%%%%%%%%%%%%%%%%%%%%%%%%%%%%%%%%%%%%%%%%%%%%%%%%%%%%%%%%%%%%%%%%%%%%%%%%%%%%%%%%%%%%%%%
Without loss of generality, we shall choose coordinates such that $k = 0, \pm 1$. Then, one
can see that the most general metric for the five-dimensional spacetime must take the
form,
\bq
\lb{2.3}
ds^{2}_{5} = g_{AB} dx^{A}dx^{B}  
 =g_{ij} dx^{i}dx^{j} - A^{2}(x^{i})d\Sigma^{2}_{k},
\eq
where $ i, j = 0, 4,\; A, B = 0, 1, 2, 3, 4$, and $d\Sigma^{2}_{k}$ is given by Eq.(\ref{2.1}). 
In such coordinates, the two 3-branes are located on the hypersurfaces
\bq
\lb{2.4}
\phi_{1}\left(x^{i}\right) = 0, \;\;\; \phi_{2}\left(x^{i}\right) = 0,
\eq
as shown by Fig. 1(a).
Clearly, the metric (\ref{2.3}) is invariant under the coordinate transformation,
\bq
\lb{2.5}
{x'}^{i} = f^{i}\left({x}^{j}\right).
\eq
As mentioned in the last section, the orbifold symmetry manifests itself explicitly
only in coordinates in which the two 3-branes are all at rest. Therefore, before imposing
such a symmetry, using  one degree of the freedom of Eq.(\ref{2.5}), we first 
map the two 3-branes to the fixed points $y = 0$ and $y = y_{c}$, for example, by 
setting 
\bq
\lb{2.6}
y = \frac{\phi^{\alpha}_{1}}{\phi^{\alpha}_{1} + \beta\phi^{\gamma}_{2}} y_{c},
\eq
where $y_{c}, \; \alpha, \; \beta$ and $\gamma$ are arbitrary constants [cf. Fig. 1(b)]. 
Without loss of generality, in this paper we assume $y_{c} > 0$ and consider the brane 
located in this surface as our TeV brane. Clearly, by properly choosing
these constants, the above coordinate transformation is non-singular \cite{NoteA}. 
Then, we can use the other degree of freedom of Eq.(\ref{2.5}) to set $g_{04} = 0$, 
so that the metric can be cast in the form,
\bq
\lb{2.7}
ds^{2}_{5} =  N^{2}\left(t,y\right)dt^{2} - A^{2}\left(t,y\right)d\Sigma^{2}_{k} 
- B^{2}\left(t,y\right)dy^{2}.
\eq

It should be noted that in \cite{BCG00} the authors used the two degrees of freedom of
Eq.(\ref{2.5}) to set $N(t,y) = B(t,y)$ and $g_{ty} = 0$ for the case of one 3-brane. 
In such chosen ($t, y$)-coordinates, the 3-brane in general is located on the hypersurface 
$y = \chi(t)$. Then, using the remaining gauge freedom, $t = t' + [\chi(t' + y')  
- \chi(t' - y')]/2$, and $y = y' + [\chi(t' + y')  + \chi(t' - y')]/2$, one can bring 
the 3-brane to the fixed point $y' = 0$, while still keep the metric in the same form
in terms of $t'$ and $y'$. For details, we refer readers to \cite{BCG00}. Clearly,
this is possible only for the case of one 3-brane. For the case of two 3-branes, 
in general we have $N(t,y) \not= B(t,y)$, as shown above.

%%%%%%%%%%%%%%%%%%%%%%%%%%%%%%%%%%%%%%%%%%%%%%%%%%%%%%%%%%%%%%%%%%%%%%%%%%%%%%%
\begin{figure}
\includegraphics[width=\columnwidth]{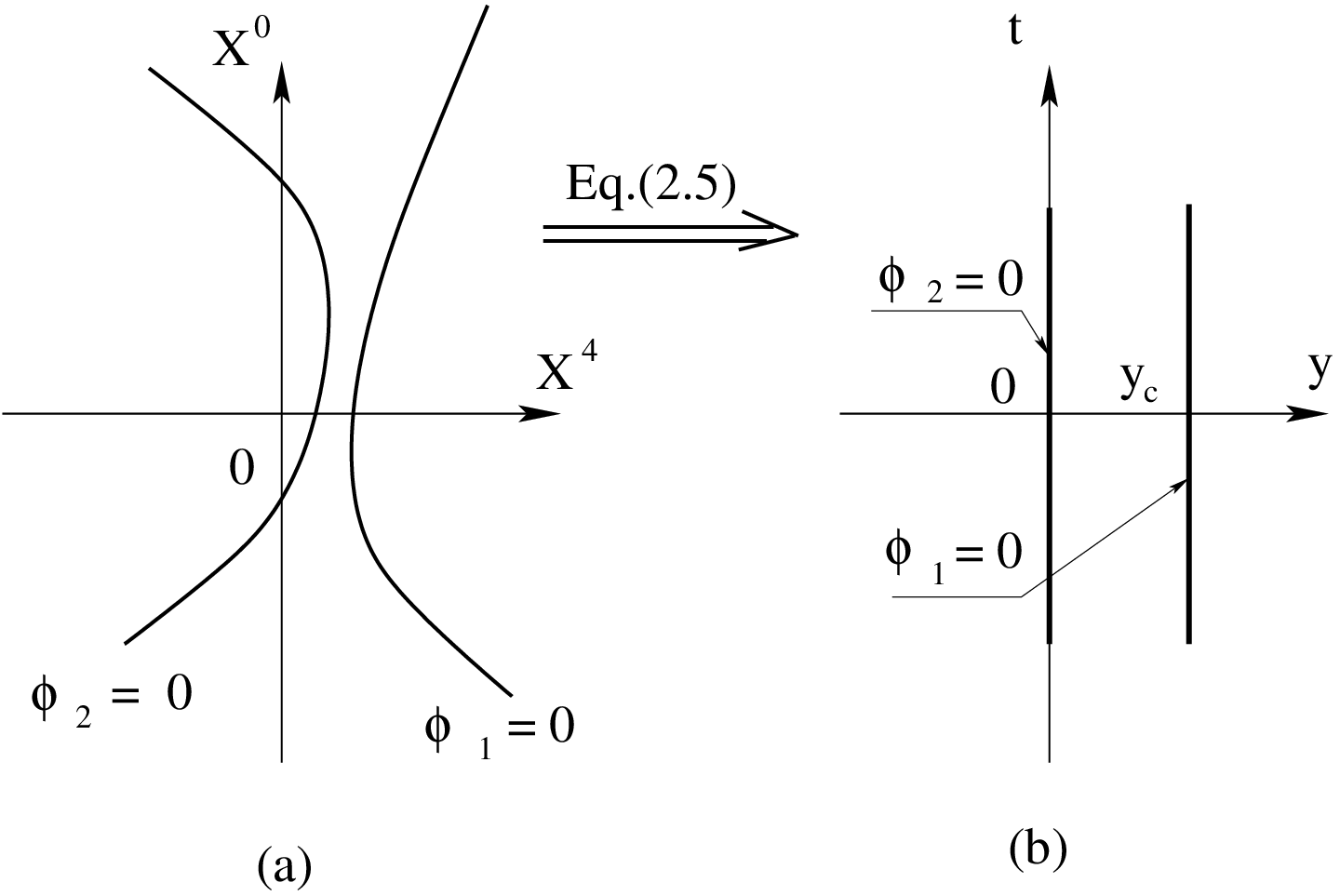}
\caption{The trajectories of two orbifold 3-branes: (a) In the ($x^{0}, x^{4}$)-plane.
(b) In the ($t, y$)-plane, in which the two 3-branes are at rest, but the proper
distance between then in general is time-dependent.} \label{fig1}
\end{figure} 
%%%%%%%%%%%%%%%%%%%%%%%%%%%%%%%%%%%%%%%%%%%%%%%%%%%%%%%%%%%%%%%%%%%%%%%%%%%%%%%

The non-vanishing components of the Einstein tensor for the metric (\ref{2.7})
 are given by
\bqn
\lb{2.7a}
G_{00}  &=& - 3\frac{N^{2}}{B^{2}}\left\{\frac{A_{,yy}}{A}
             + \frac{A_{,y}}{A}\left(\frac{A_{,y}}{A} - \frac{B_{,y}}{B}\right)\right\}\nb\\
	     & & + 3 \frac{A_{,t}}{A}\left(\frac{A_{,t}}{A} + \frac{B_{,t}}{B}\right)
	          +  3k \frac{N^{2}}{A^{2}},\nb\\
G_{04}  &=& - 3 \left\{\frac{A_{,ty}}{A}
             - \frac{A_{,t}}{A} \frac{N_{,y}}{N} 
	     - \frac{B_{,t}}{B} \frac{A_{,y}}{A}\right\},\nb\\ 
G_{ab}  &=& \frac{A^{2}}{B^{2}}\gamma_{ab}\left\{2\frac{A_{,yy}}{A} + \frac{N_{,yy}}{N}
        + \frac{A_{,y}}{A}\left(\frac{A_{,y}}{A} - 2\frac{B_{,y}}{B}\right)\right.\nb\\
       & &  \left.+ \frac{N_{,y}}{N}\left(2\frac{A_{,y}}{A} - \frac{B_{,y}}{B}\right)\right\}\nb\\
       & & - \frac{A^{2}}{N^{2}}\gamma_{ab}\left\{2\frac{A_{,tt}}{A} + \frac{B_{,tt}}{B} 
         + \frac{A_{,t}}{A}\left(\frac{A_{,t}}{A} + 2\frac{B_{,t}}{B}\right)\right.\nb\\
       & &  \left.- \frac{N_{,t}}{N}\left(2\frac{A_{,t}}{A} + \frac{B_{,t}}{B}\right)\right\}
            - k\gamma_{ab},\nb\\
G_{44}  &=& - 3\frac{B^{2}}{N^{2}}\left\{\frac{A_{,tt}}{A}
             + \frac{A_{,t}}{A}\left(\frac{A_{,t}}{A} - \frac{N_{,t}}{N}\right)\right\}\nb\\
	     & & + 3 \frac{A_{,y}}{A}\left(\frac{A_{,y}}{A} + \frac{N_{,y}}{N}\right)
	         -  3 k \frac{B^{2}}{A^{2}},
\eqn
where   $F_{,y} \equiv \partial F/\partial y$, etc.

The reflection symmetry with respect to the two branes can be obtained by the replacement
\bq
\lb{2.8}
 y \rightarrow \left|y\right|, 
\eq
so that the most general metric for two 3-branes finally takes the form,
\bq
\lb{2.7b}
ds^{2}_{5} =  N^{2}\left(t,\left|y\right|\right)dt^{2} 
- A^{2}\left(t,\left|y\right|\right)d\Sigma^{2}_{k} 
- B^{2}\left(t,\left|y\right|\right)dy^{2},
\eq
where
\bqn 
\lb{2.9}
\left|y\right| & \equiv& \cases{ ...&, ..., \cr
y + 2y_{c}&, $ - 2y_{c} \le y \le - y_{c}$,\cr
- y  &, $ - y_{c} \le y \le 0$,\cr
 y  &, $0 \le y \le y_{c}$,\cr
 2y_{c}- y  &, $  y_{c} \le y \le 2y_{c}$,\cr
 ...&, ...,\cr}\nb\\
&=& \sum^{\infty}_{n = -\infty}{\left(2ny_{c} -y\right)
  H\left(y - \left(2n-1\right)y_{c}\right)}\nb\\
& & \;\;\;\;\;\;\;\;\;\;\; \times
\left[1 - H\left(y - 2ny_{c}\right)\right]\nb\\
& & 
+ \sum^{\infty}_{n = -\infty}{\left(y - 2ny_{c}\right)H\left(y - 2ny_{c}\right)}\nb\\
& & \;\;\;\;\;\;\;\;\;\;\;\; \times
\left[1 - H\left(y - \left(2n+1\right)y_{c}\right)\right],
\eqn 
as shown in Fig. 2, where $H(x)$ denotes the Heavside function, defined by,
\bq
\lb{2.12a}
H(x) = \cases{1, & $x > 0$,\cr
              0, & $ x < 0$.\cr}
\eq
Hence, we obtain
\bqn
\lb{2.9aa}
\frac{d|y|}{dy} &=&  \sum^{\infty}_{n = -\infty}{-
  H\left(y - \left(2n-1\right)y_{c}\right) }\nb\\
& & \;\;\;\;\;\;\;\;\;\;\;
\times \left[1 - H\left(y - 2ny_{c}\right)\right]\nb\\
& & 
+ \sum^{\infty}_{n = -\infty}{ H\left(y - 2ny_{c}\right)}\nb\\
& & \;\;\;\;\;\;\;\;\;\;\;\;
\times
\left[1 - H\left(y - \left(2n+1\right)y_{c}\right)\right],\nb\\
\frac{d^{2}|y|}{dy^{2}} &=&  2\sum^{\infty}_{n = -\infty}{\delta(y - 2ny_{c})}\nb\\
& & 
- 2 \sum^{\infty}_{n = -\infty}{\delta\left(y - (2n+1)y_{c}\right)},
\eqn 
where $\delta(x)$ denotes the Dirac function,  with
\bqn
\lb{2.12b}
\frac{dH(x)}{dx} &=& \delta(x),\nb\\
\int^{\infty}_{-\infty}{f(x) \delta(x)dx} &=& f(0),
\eqn
for any given test function $f(x)$.
%%%%%%%%%%%%%%%%%%%%%%%%%%%%%%%%%%%%%%%%%%%%%%%%%%%%%%%%%%%%%%%%%%%%%%%%%%%%%%%
\begin{figure}
\includegraphics[width=\columnwidth]{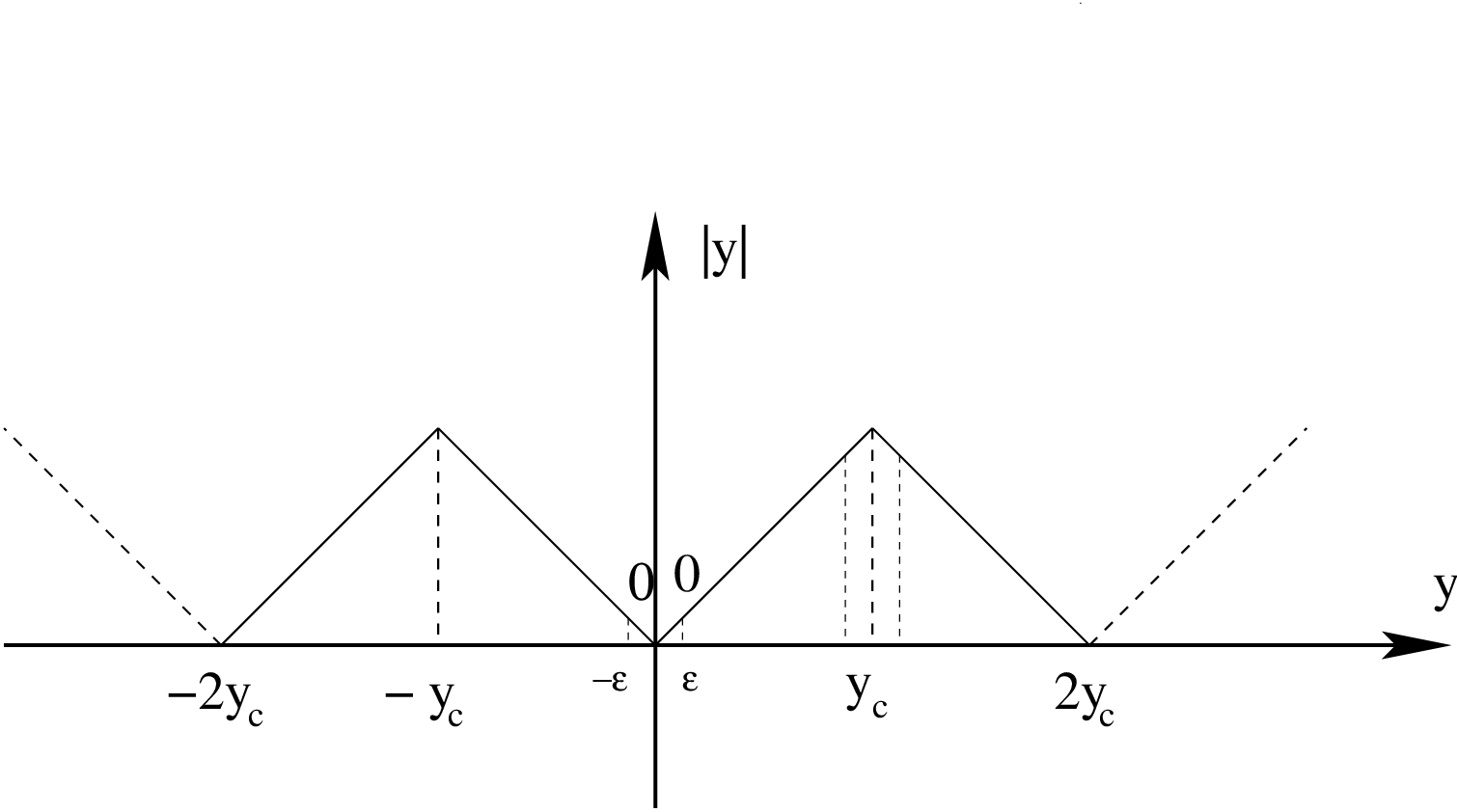}
\caption{The function $\left|y\right|$  defined by Eq.(\ref{2.9}). }
\label{fig2}
\end{figure} 
%%%%%%%%%%%%%%%%%%%%%%%%%%%%%%%%%%%%%%%%%%%%%%%%%%%%%%%%%%%%%%%%%%%%%%%%%%%%%%%

However, the orbifold symmetry first restricts  $y$ to the range, $ - y_{c} \le y
\le y_{c}$, and then identifies the points $\left(x^{\mu}, \; y\right)$ with the 
ones $\left(x^{\mu}, \; - y\right)$, so that $y$ finally takes its values only in 
the range, $0 \le y \le y_{c}$. Then, Eq.(\ref{2.9aa}) yields
\bqn 
\lb{2.9ab}
\frac{d|y|}{dy} &=&  1, \nb\\ 
\frac{d^{2}|y|}{dy^{2}} &=&  2\delta(y) 
 - 2  \delta\left(y - y_{c}\right),
\eqn 
for $y \in [0, y_{c}]$.

It should be noted that, although the two 3-branes
are all at rest in the above chosen gauge, which will be referred to as {\em the 
canonical gauge},  the proper distance between them in general is time-dependent, 
and   given by
\bq
\lb{2.9a}
{\cal{D}}(t) = \int^{y_{c}}_{0}{B(t,y)dy}.
\eq 
The moment where ${\cal{D}}(t) = 0$ is the one when the two 3-branes collide.

We also note that the expressions given by Eq.(\ref{2.7a}) for the Einstein 
tensor hold even for spacetimes with orbifold symmetry, 
but now in the sense of distribution. To see this, let us consider a given 
function $F(t, |y|)$, for which we find  
\bqn
\lb{2.10}
\frac{\partial F(t, |y|)}{\partial t} &=&  
\frac{\partial F(t,y)}{\partial t},\nb\\
\frac{\partial F(t, |y|)}{\partial y} &=&  
\frac{\partial F(t,y)}{\partial y},\nb\\
\frac{\partial^{2} F(t, |y|)}{\partial t\partial y} &=&  
\frac{\partial^{2} F(t,y)}{\partial t\partial y},\nb\\
\frac{\partial^{2} F(t, |y|)}{\partial t^{2}} &=&  
\frac{\partial^{2} F(t,y)}{\partial t^{2}},\nb\\
\frac{\partial^{2} F(t, |y|)}{\partial y^{2}} &=& 
  \frac{\partial^{2} F(t,y)}{\partial y^{2}} 
     + \left[F_{,y}\right]_{0} \delta(y) \nb\\
& & + \left[F_{,y}\right]_{c} \delta\left(y-y_{c}\right), 
\eqn
 for $y \in \left[0, y_{c}\right]$,  where
\bqn
\lb{2.11}
\left[F_{,y}\right]_{0} &\equiv& \lim_{y \rightarrow 0^{+}}
     {\frac{\partial F(t,y)}{\partial y}}
      - \lim_{y \rightarrow 0^{-}}{\frac{\partial F(t,-y)}{\partial y}}\nb\\
      &=& 2 \left.\frac{\partial F(t,y)}{\partial y}\right|_{y=0^{+}},\nb\\
\left[F_{,y}\right]_{{c}} &\equiv& \lim_{y \rightarrow y_{c}^{+}}
     {\frac{\partial F(t,2y_{c}-y)}{\partial y}}
      - \lim_{y \rightarrow y_{c}^{-}}{\frac{\partial F(t,y)}{\partial y}}
\nb\\
      &=& -2 \left. \frac{\partial F(t,y)}{\partial y}\right|_{y=y_{c}^{-}}.
\eqn
Note that in writing Eq.(\ref{2.11}) we had assumed that the derivatives of $F(t, |y|)$
exist from each side of the limits. It can also be shown that 
\bqn
\lb{2.13}
& & H^{m}(x) = H(x),\;\;\; \left[1 - H(x)\right]^{m} = 1 - H(x),\nb\\
& &  H(x)\left[1 - H(x)\right] = 0, \; H(x)\delta(x) = \frac{1}{2}\delta(x),\nb\\
& & \left[F^{+}(x) H(x) + F^{-}(x) \left(1- H(x)\right)\right] \delta(x) \nb\\
& & \;\;\;\;\;\;\;\;\; = \frac{1}{2} \left[F^{+}(0)   + F^{-}(0)\right] \delta(x),
\eqn
in the  sense of distributions, where $m$ is an integer, and
\bqn
\lb{2.13a}
F^{+}(0) &\equiv& \lim_{x \rightarrow 0^{+}}{F^{+}(x)},\nb\\
F^{-}(0) &\equiv& \lim_{x \rightarrow 0^{-}}{F^{-}(x)}.
\eqn
Inserting Eqs.(\ref{2.10}) - (\ref{2.13a}) into Eq.(\ref{2.7a}),
we find that the Einstein tensor  corresponding to the metric (\ref{2.7b}) 
can be written as  
\bq
\lb{2.14}
G_{AB} = G_{AB}^{(D)}
 + G_{AB}^{(0)}\delta(y) +  G_{AB}^{(c)}\delta\left(y-y_{c}\right),
\eq
for $y \in [0, y_{c}]$,
where $G_{AB}^{(D)}$ denotes the Einstein tensor calculated in the region
$y \in \left(0, y_{c}\right)$ and given by Eq.(\ref{2.7a}),  and 
\bqn
\lb{2.15a}
G_{AB}^{(0)} &=& \frac{A^{2}_{0}}{B^{2}_{0}}\left(2\frac{\left[A_{,y}\right]_{0}}{A_{0}}
    + \frac{\left[N_{,y}\right]_{0}}{N_{0}}\right) 
   \gamma_{ab}\delta^{a}_{A} \delta^{b}_{B}\nb\\
& & - 3 \frac{N^{2}_{0}}{B^{2}_{0}}\frac{\left[A_{,y}\right]_{0}}{A_{0}}
\delta^{0}_{A}\delta^{0}_{B},\\
\lb{2.15b}
 G_{AB}^{(c)} &=& \frac{A^{2}_{c}}{B^{2}_{c}}\left(2\frac{\left[A_{,y}\right]_{c}}{A_{c}}
    + \frac{\left[N_{,y}\right]_{c}}{N_{c}}\right) 
    \gamma_{ab}\delta^{a}_{A} \delta^{b}_{B}\nb\\
& & - 3 \frac{N^{2}_{c}}{B^{2}_{c}}\frac{\left[A_{,y}\right]_{c}}{A_{c}}
\delta^{0}_{A}\delta^{0}_{B},
\eqn
with
\bq
\lb{2.16}
F_{0}(t)   \equiv F\left(t, 0\right), \;\;\; F_{c} (t)  \equiv F\left(t, y_{c}\right),
\eq
where $F \equiv \left\{N, A, B\right\}$. Assuming that the energy-momentum tensor (EMT)
takes the form,
\bqn
\lb{2.17}
T_{AB} &=& T_{AB}^{(D)} + \frac{1}{B_{0}(t)}T_{AB}^{(0)}\delta(y) \nb\\
& & + \frac{1}{B_{c}(t)} T_{AB}^{(c)}\delta\left(y-y_{c}\right),
\eqn
where $T_{AB}^{(D)}$ denotes the bulk EMT, and $T_{AB}^{(0)}$ and $T_{AB}^{(c)}$
the EMT's of the two 3-branes,   the Einstein equations
take the form,
\bqn
\lb{2.18a}
G_{AB}^{(D)} &=& \kappa_{5}T_{AB}^{(D)} + \Lambda g_{AB}, 
\;\; y\in\left(0, y_{c}\right),\\
\lb{2.18b}
G_{\mu\nu}^{(0)} &=& \frac{1}{B_{0}(t)} \left(\kappa_{5}T_{\mu\nu}^{(0)} 
+ \lambda_{0} g_{\mu\nu}^{(0)}\right), 
\;\; (y = 0),\\
\lb{2.18c}
G_{\mu\nu}^{(c)} &=& \frac{1}{B_{c}(t)} \left(\kappa_{5}T_{\mu\nu}^{(c)} 
+ \lambda_{c} g_{\mu\nu}^{(c)}\right), 
\;\; (y = y_{c}),
\eqn
where     
\bq
\lb{2.19}
g_{\mu\nu}^{(0)} (t) \equiv g_{\mu\nu}(t, 0),\;\;\;
g_{\mu\nu}^{(c)}(t) \equiv g_{\mu\nu}(t, y_{c}).
\eq
The constants $\Lambda,\; \lambda_{0}$ and $ \lambda_{c}$ denote, 
respectively, the bulk and the brane 
cosmological constants. Assuming that the EMT's of the two 
3-branes are given by perfect fluids,
\bqn
\lb{2.20}
T^{(0)}_{\mu\nu} &=& \left(\rho_{0} + p_{0}\right) u^{(0)}_{\mu}u^{(0)}_{\nu}
- p_{0}g_{\mu\nu}^{(0)},\nb\\
T^{(c)}_{\mu\nu} &=& \left(\rho_{c} + p_{c}\right) u^{(c)}_{\mu}u^{(c)}_{\nu}
- p_{c} g_{\mu\nu}^{(c)},
\eqn
where
\bq
\lb{2.21}
u^{(0)}_{\mu} = N_{0}\delta^{0}_{\mu}, \;\;\; u^{(c)}_{\mu} = N_{c}\delta^{0}_{\mu},
\eq
we find that Eqs.(\ref{2.18b}) and (\ref{2.18c}) reduce to
\bqn
\lb{2.22}		
\frac{1}{B_{0}}\left(2\frac{\left[A_{,y}\right]_{0}}{A_{0}}
              + \frac{\left[N_{,y}\right]_{0}}{N_{0}}\right) 
                + \lambda_{0} &=& \kappa_{5} p_{0},\nb\\
\frac{3}{B_{0}}\frac{\left[A_{,y}\right]_{0}}{A_{0}}
                + \lambda_{0} &=& - \kappa_{5} \rho_{0},			
\eqn
and
\bqn
\lb{2.23}
 \frac{1}{B_{c}}\left(2\frac{\left[A_{,y}\right]_{c}}{A_{c}}
              + \frac{\left[N_{,y}\right]_{c}}{N_{c}}\right) 
                + \lambda_{c} &=& \kappa_{5} p_{c},\nb\\
\frac{3}{B_{c}}\frac{\left[A_{,y}\right]_{c}}{A_{c}}
                + \lambda_{c} &=& - \kappa_{5} \rho_{c}.
\eqn				

This completes our general description of two homogeneous and 
isotropic 3-branes with orbifold symmetry, in which the bulk can be filled
with any kind of matter fields. In the following, we consider some specific 
solutions.

\section{Dynamical Two 3-Branes in $AdS_{5}$}

\renewcommand{\theequation}{3.\arabic{equation}}
\setcounter{equation}{0}  

In this section, we first consider the global properties of $AdS_{5}$, and 
then apply the formulas developed in the last section to study two 3-branes
in the background of $AdS_{5}$. 

\subsection{$AdS_{5}$ Space in Horospheric Coordinates}

$AdS_{5}$ can be realized as a hyperboloid embedded in a flat six-dimensional
space with two timelike coordinates,
\bq
\lb{eq3.1}
\eta_{AB} Y^{A}Y^{B} = {\ell}^{-2},\; (A,\; B = 0, 1, ..., 5),
\eq
where $\eta_{AB} = diag \{+ 1, - 1, - 1, - 1, - 1, + 1\}$, and ${\ell} \equiv 
\left(-\Lambda/6\right)^{1/2}$, where $\Lambda$ denotes the cosmological
constant of the five-dimensional bulk. In fact, introducing the Einstein
universe (EU) coordinates, $\{t_{c}, \xi, \theta, \phi, \psi\}$, via the 
relations,
\bqn
\lb{eq3.2}
{\ell} Y^{0} &=& \cos t_{c}\sec\xi,\nb\\
{\ell} Y^{1} &=& \sin\theta\sin\phi\cos\psi\tan\xi,\nb\\
{\ell} Y^{2} &=& \sin\theta\sin\phi\sin\psi\tan\xi,\nb\\
{\ell} Y^{3} &=& \sin\theta\cos\phi\tan\xi,\nb\\
{\ell} Y^{4} &=& \cos\theta\tan\xi,\nb\\
{\ell} Y^{5} &=& - \sin t_{c}\sec\xi,
\eqn
we find that
\bqn
\lb{eq3.3}
ds^{2}_{5} &=& \left.\eta_{AB} dY^{A} d Y^{B}\right|_{hyperboloid}\nb\\
&=& \frac{1}{\left({\ell}\cos\xi\right)^{2}}
\left(dt^{2}_{c} - d\xi^{2} -\sin\xi^{2}d\Omega_{3}^{2}\right)\nb\\
&\equiv& \frac{1}{\left({\ell}\cos\xi\right)^{2}}ds^{2}_{5(E)},
\eqn
where $d\Omega_{3}^{2} [\equiv d\theta^{2} + \sin\theta^{2} d\phi^{2}
+ \sin\theta^{2} \sin\phi^{2} d\psi^{2}]$ is the line element on the
three-sphere $S^{3}$. The range of the EU coordinates that covers the
whole  $AdS_{5}$ space is $- \pi \le t_{c} \le \pi$, $0 \le \xi \le \pi/2$,
$0 \le \theta, \; \phi \le \pi$, and $ 0 \le \psi \le 2\pi$. The timelike
hypersurface $\xi = \pi/2$ represents the spatial infinite. The spacelike
hypersurfaces $t_{c} = \pm\pi$ are identified, which represent closed 
timelike curves (CTC's). The topology of $AdS_{5}$ is $S^{1}$ (time) $\times
R^{4}$ (space). To avoid CTC's for $AdS_{5}$, one may simply unwrap the 
timelike coordinate $t_{c}$ and extend it to the range, $ - \infty < t_{c}
< \infty$, so the resulting space has the topology $R^{1}$ (time) $\times
R^{4}$ (space), which will be referred to as $CAdS_{5}$. Since the
spatial infinity is timelike, neither $CAdS_{5}$ nor $AdS_{5}$ has Cauchy
surface. In the rest of this paper, for the sake of simplicity, when we talk 
about $AdS_{5}$, we always means $CAdS_{5}$. 

It is interesting to note that $CAdS_{5}$ covers only half of the static 
Einstein universe given by $ds^{2}_{5(E)}$  in Eq.(\ref{eq3.3}) with the same 
range for $t_{c}, \theta, \phi$ and $ \psi$, but now $0 \le \xi \le \pi$ 
[cf. Fig. 3].  

%%%%%%%%%%%%%%%%%%%%%%%%%%%%%%%%%%%%%%%%%%%%%%%%%%%%%%%%%%%%%%%%%%%%%%%%%%%%%%%
\begin{figure}
\includegraphics[width=\columnwidth]{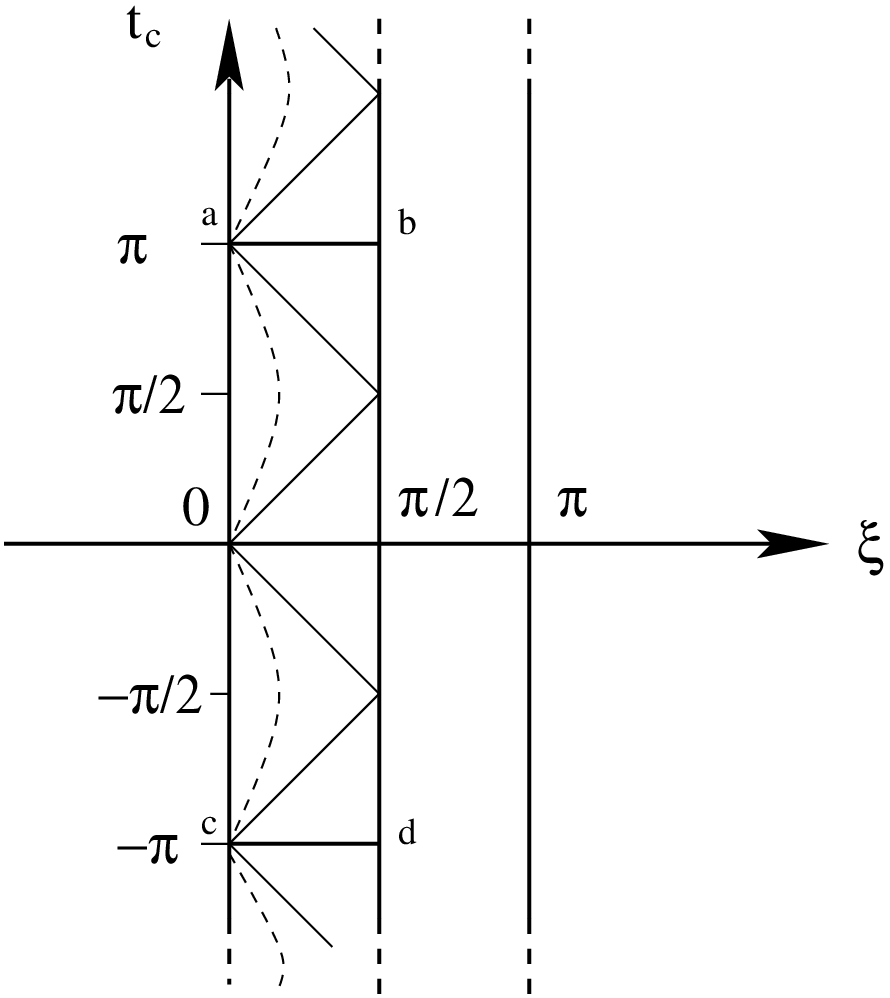}
\caption{The $AdS_{5}$ space seen on the Einstein cylinder. The coordinates
$\theta, \; \phi$ and $\psi$ are suppressed. Thus, each point represents a 
three-sphere. The lines $ab$ and $cd$ are identified, which 
represent CTC's. Unwrapping  $t_{c}$ coordinate  and then extending it
to the range, $ t_{c} \in (-\infty, \infty)$, one can avoid CTC's.  
The dotted periodic lines represent timelike geodesics, and the $45^{o}$ 
straight lines the null geodesics. The vertical line $\xi = \pi/2$ 
represents the spatial infinity, and a timelike geodesic never reaches it.}  
\label{fig3} 
\end{figure} 
%%%%%%%%%%%%%%%%%%%%%%%%%%%%%%%%%%%%%%%%%%%%%%%%%%%%%%%%%%%%%%%%%%%%%%%%%%%%%%%

To study RS1 model, it is found convenient to introduce the horospheric 
coordinates, $\{t, x^{1}, x^{2}, x^{3}, z\}$, via the relations,
\bqn
\lb{eq3.4}
& & {\ell} Y^{0} = -t/z,\nb\\
& & {\ell} Y^{i} = x^{i}/z,  \nb\\
%& & {\ell} Y^{2} = x^{2}/z,  \nb\\
%& & {\ell} Y^{3} = x^{3}/z,  \nb\\
& & {\ell}\left(Y^{5} + Y^{4}\right) = 1/({\ell}z),\nb\\
& & {\ell}\left(Y^{5} - Y^{4}\right) = {\ell}z 
   + {\ell}^{2}\left({\bf x}^{2} - t^{2}\right)/z,
\eqn
where $ - \infty < t, x^{i}, z < \infty \; (i = 1, 2, 3)$.
Reversing the sign of $z$ in the above expressions reverses 
the signs of all $Y^{A}$'s, which corresponds to the antipodal
map on the $AdS_{5}$ hyperboloid \cite{Gib93}. Thus, in order
to cover all of $AdS_{5}$ space, at least two horospheric charts,
one with positive $z$ and the other with negative $z$ are needed.
Vanishing $z$ corresponds to the spacelike infinity of $AdS_{5}$
($\xi = \pi/2$). In terms of the horospheric coordinates, the 
$AdS_{5}$ metric takes the conformally-flat form,  
\bq
\lb{eq3.5}
ds^{2}_{5} = \frac{1}{\left(\ell z\right)^{2}}
\left(dt^{2} - dz^{2} - d\Sigma^{2}_{0}\right).
\eq
As Gibbons first noticed \cite{Gib93}, although the distance alone 
the spacelike geodesics of constant $t$ and $x^{i}$ diverges as $|z| 
\rightarrow \infty$, the spacetime described by the metric 
(\ref{eq3.5}) is not geodesically complete with respect to timelike 
and null geodesics. Thus, to obtain a geodesically complete $AdS_{5}$
space, one needs to extend the region covered by metric (\ref{eq3.5}). 
Since the metric has Poincar\'e invariance on the hypersurfaces $z = $ 
constant, we may suppress the coordinates $x^{i}$, and only consider the 
extension of the $(1+1)$-dimensional $AdS_{2}$ space,
\bq
\lb{eq3.6}
ds^{2}_{2} = \frac{1}{\left(\ell z\right)^{2}}
\left(dt^{2} - dz^{2}\right).
\eq
Indeed,  understanding the causal structure of 
$AdS_{n}\; (n \ge 2)$ reduces to that of $AdS_{2}$  \cite{CGS93}. Clearly, metric 
(\ref{eq3.6}) is conformally related to the ($1+1$)-dimensional Minkowski 
spacetime, $ds^{2}_{M_{2}} = dt^{2} - dz^{2}$.
Then, the corresponding Penrose diagram is given by Fig. 4. 
%%%%%%%%%%%%%%%%%%%%%%%%%%%%%%%%%%%%%%%%%%%%%%%%%%%%%%%%%%%%%%%%%%%%%%%%%%%%%%%
\begin{figure}
\includegraphics[width=\columnwidth]{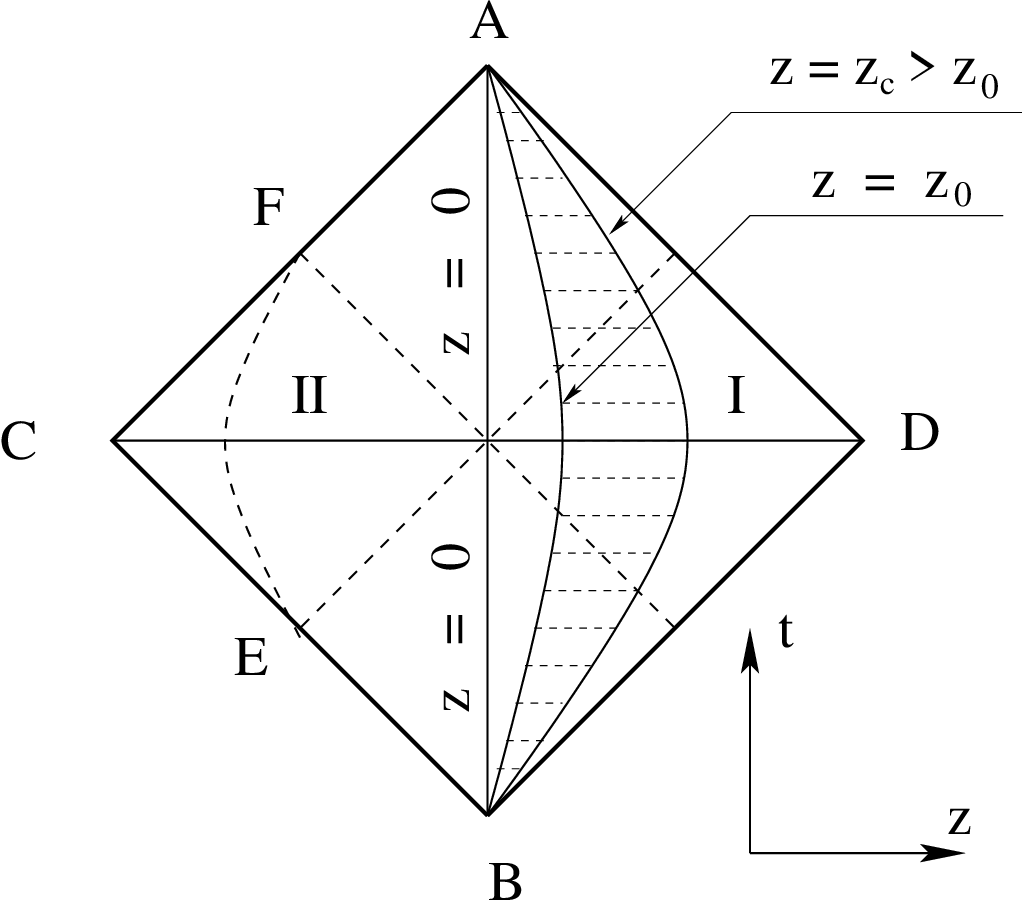}
\caption{The Penrose diagram of $AdS_{2}$ space 
defined by metric (\ref{eq3.6}). The vertical curves are $z = z_{0}$ and
$z = z_{c}$, respectively, with $z_{c} > z_{0} > 0$.}  
\label{fig4} 
\end{figure} 
%%%%%%%%%%%%%%%%%%%%%%%%%%%%%%%%%%%%%%%%%%%%%%%%%%%%%%%%%%%%%%%%%%%%%%%%%%%%%%%
Points on each side of the vertical line $AB$   are related by 
the anti-podal map, $z \rightarrow - z$. The dashed curve
$EF$ is a timelike geodesics, which reaches the null boundary $AC$ within a
finite proper time. Thus, extension beyond it is needed, so does across the 
other three null boundaries. However, these boundaries are Cauchy horizons.
For example, events beyond the boundaries $AC$ and $AD$ are not uniquely 
determined by initial data  imposed on the line $CD$. As a result, such
extension is not unique.  As  noticed by Gibbons, the coordinate $z$ 
jumps from infinitely large positive values to infinitely large negative values
as one crosses the horizons \cite{Gib93}. Therefore, one possible extension 
is given by Fig. 5, which covers the whole plane with infinite lattice
of diamonds  \cite{CGS93}.

%%%%%%%%%%%%%%%%%%%%%%%%%%%%%%%%%%%%%%%%%%%%%%%%%%%%%%%%%%%%%%%%%%%%%%%%%%%%%%%
\begin{figure}
\includegraphics[width=\columnwidth]{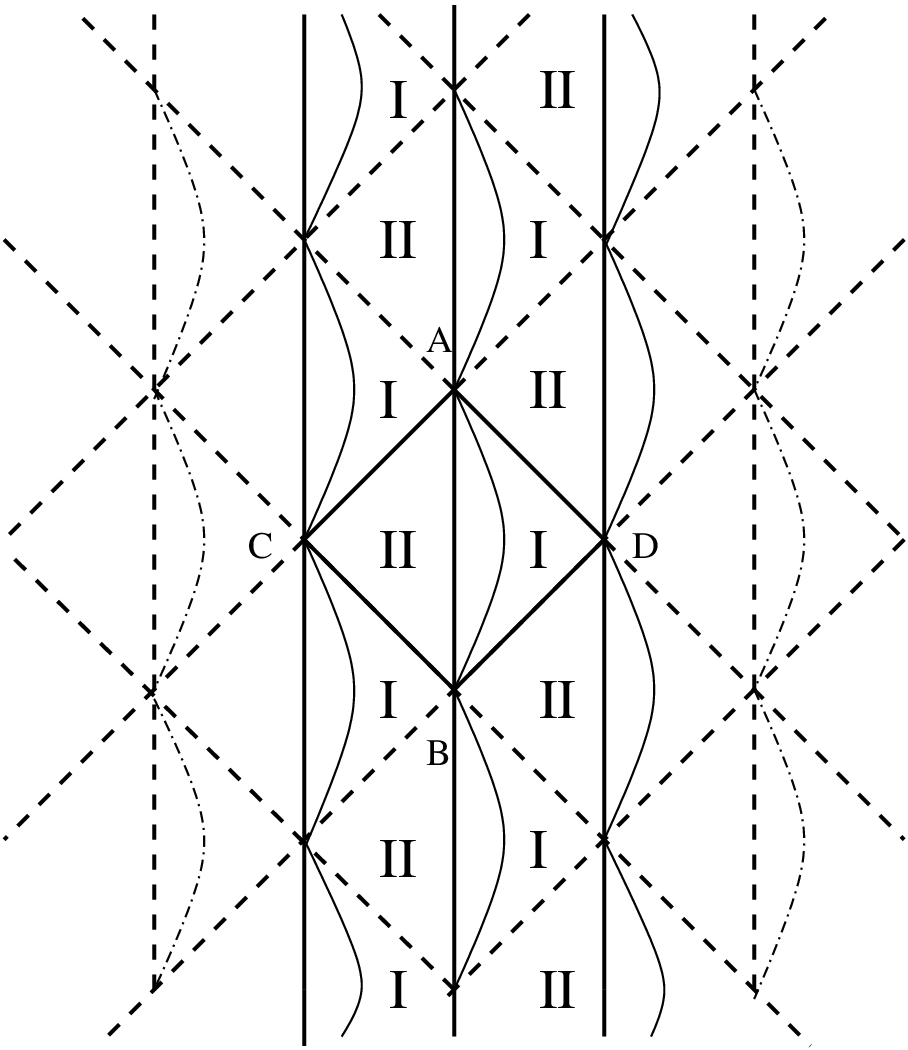}
\caption{The Penrose diagram of  the extended $AdS_{2}$ space. The vertical
curves,  both solid and dashed,  represent hypersurfaces of constant $z > 0$.}  
\label{fig5} 
\end{figure} 
%%%%%%%%%%%%%%%%%%%%%%%%%%%%%%%%%%%%%%%%%%%%%%%%%%%%%%%%%%%%%%%%%%%%%%%%%%%%%%%

\subsection{RS1 Model}

To obtain the RS1 solution, we first make the coordinate transformation,
\bq
\lb{eq3.7}
 {\ell} z = e^{\ell y},
\eq
so that metric (\ref{eq3.5}) takes the form,
\bq
\lb{eq3.8}
ds^{2}_{5} = e^{- 2\ell y} \left(dt^{2}  - d\Sigma^{2}_{0}\right)
- dy^{2}.
\eq
Then, applying the replacement (\ref{2.8}) in the above metric coefficients, we obtain
the RS metric (\ref{1.5}) with $y = r_{c}\phi$. Comparing it with the general form
of the metric Eq.(\ref{2.7b}), we find  
\bqn
\lb{3.2}
N\left(t, |y|\right) &=& A\left(t, |y|\right) = e^{-{\ell}|y|}, \nb\\
B\left(t, |y|\right) &=& 1, \;\;\; k = 0.
\eqn
Substituting Eq.(\ref{3.2}) into Eqs.(\ref{2.22}) and (\ref{2.23}), we obtain
\bqn
\lb{3.3}
\rho_{0} &=& - p_{0} = \frac{1}{\kappa_{5}} \left(6{\ell} - \lambda_{0}\right),\nb\\
\rho_{c} &=& - p_{c} = -\frac{1}{\kappa_{5}} \left(6{\ell} + \lambda_{c}\right).
\eqn
RS chose 
\bq
\lb{3.4}
\lambda_{0} = -  \lambda_{c} = 6{\ell},
\eq
so that no matter fields appear on the two   branes. In Fig. 4,   
the two 3-branes are represented by the two solid 
vertical curves with $z_{0} \equiv \ell^{-1}$
and $z_{c} \equiv \ell^{-1} e^{\ell y_{c}}$, respectively. 
The bulk is the region between these two curves, the shaded region. 
As shown in the last sub-section, the spacetime is not geodesically complete, 
and extension is needed. One possible extension is that given by Fig. 5
but with two vertical curves, one is $z = z_{0}$ and the other is $z = z_{c}$, 
where the bulk is  the region between the two curves. Clearly, in this extension 
we have infinite number of RS1 universes.

\subsection{Dynamical 3-branes in $AdS_{5}$}

To find new solutions, we first make the coordinate transformation,
\bqn
\lb{3.5a}
t &=&  F\left(t'+f(y')\right) + G\left(t'-f(y')\right), \nb\\
z &=& F\left(t'+f(y')\right) - G\left(t'-f(y')\right), 
\eqn
so that  metric (\ref{eq3.5}) takes the form,
\bq
\lb{3.5}
ds^{2}_{5} = \frac{4F'G'}{{\ell}^{2}(F-G)^{2}}\left(d{t}^{2} 
- {f'}^{2}(y)d{y}^{2}\right) - 
\frac{1}{{\ell}^{2}(F-G)^{2}}d\Sigma_{0}^{2},
\eq
where  $F, \; G$ and $f$ are arbitrary functions, and
a prime denotes ordinary differentiation with respect to their 
indicated arguments. Note that in writing Eq.(\ref{3.5}), we had 
dropped the primes from $t'$ and $y'$.
Replacing $y$ by $|y|$ in the metric coefficients, the resulting 
metric finally takes the form of Eq.(\ref{2.7b}), but now with
\bqn
\lb{3.6}
N(t, |y|) &=&   \frac{2\sqrt{F' G'}}
{{\ell}(F -G)},\nb\\
B(t, |y|) &=&   \frac{2\sqrt{F' G'}}
{{\ell}(F -G)} {f'}(|y|),\nb\\
A(t, |y|) &=&  \frac{1}{{\ell}(F -G)},
\eqn
where
\bq
\lb{3.6a}
F = F\left(t+f(|y|)\right),\;\;\;
G= G\left(t-f(|y|)\right), 
\eq
and $|y|$ is given by Eq.(\ref{2.9}). On the hypersurface $y = 0$, 
the metric reduces to
\bq
\lb{3.7}
\left. ds^{2}_{5}\right|_{y=0} = d\eta^{2} - a^{2}_{0}(\eta) d\Sigma_{0}^{2},
\eq
with
\bqn
\lb{3.8}
d\eta &\equiv& \frac{2\sqrt{{F'}_{0}{G'}_{0}}}{{\ell}(F_{0}-G_{0})}dt, \nb\\
F_{0} &\equiv&  F(t + f(0)),\;\;\;
G_{0} \equiv  G(t - f(0)), \nb\\
a_{0}(\eta) &\equiv& \frac{1}{{\ell}(F_{0}-G_{0})}, \;\;\;
b_{0}(\eta) \equiv \frac{1}{{\ell}(F_{0}+ G_{0})}. 
\eqn
From the above expressions it can be shown that 
\bqn
\lb{3.9}
\left. \frac{{F'}}{\sqrt{{F'}{G'}}}\right|_{y = 0}  &=& - \frac{a_{0}}{{\ell}}
\left(\frac{\dot{b}_{0}}{b_{0}^{2}} +  \frac{\dot{a}_{0}}{a_{0}^{2}}\right),\nb\\
\left. \frac{{G'}}{\sqrt{{F'}{G'}}}\right|_{y = 0}  &=& - \frac{a_{0}}{{\ell}}
\left(\frac{\dot{b}_{0}}{b_{0}^{2}} -  \frac{\dot{a}_{0}}{a_{0}^{2}}\right), 
\eqn
from which we find that
\bqn
\lb{3.10}
\frac{\dot{b}_{0}}{b_{0}^{2}} &=& 
\frac{\epsilon_{0}}{a_{0}}\left(\ell^{2} + H^{2}_{0}\right)^{1/2},\nb\\
H_{0} &\equiv& \frac{\dot{a}_{0}}{a_{0}}, \;\;\; \epsilon_{0} = \pm 1,
\eqn
where $\dot{a}_{0} \equiv da_{0}/d\eta$.  Inserting the above expressions into 
Eq.(\ref{3.9}) we find that, in order for ${F'}_{0}$ and ${G'}_{0}$ to be real, 
$\epsilon_{0}$ must be chosen such that,
\bq
\lb{3.10a}
\epsilon_{0} =\cases{ -1, &  ${F'}_{0} > 0,\; {G'}_{0} > 0$,\cr
+ 1, &  ${F'}_{0} < 0,\; {G'}_{0} < 0$.\cr}
\eq
 
Substituting Eqs.(\ref{3.6}) - (\ref{3.10a}) into Eq.(\ref{2.22}) we obtain
\bqn
\lb{3.11a}
\left(H_{0}^{2} + \ell^{2}\right)^{1/2} &=& -\frac{1}{6}
\kappa_{5}\epsilon_{0}\epsilon^{0}_{f}
   \left({\rho}_{0} + \bar{\lambda}_{0}\right),\\
\lb{3.11b}
\frac{\dot{H}_{0}}{\left(H_{0}^{2} + \ell^{2}\right)^{1/2}} &=&  
 \frac{1}{2} \kappa_{5}\epsilon_{0}\epsilon^{0}_{f}
\left({\rho}_{0} + {p}_{0}\right),
\eqn
where $ \bar{\lambda}_{0} \equiv \lambda_{0}/\kappa_{5}$, 
and $\epsilon^{0}_{f} \equiv f'(0)/\left|f'(0)\right|$.

Similarly, on the TeV brane $y = y_{c}$,   introducing the quantities,
\bqn
\lb{3.13}
d\tau &\equiv& \frac{2\sqrt{{F'}_{c}{G'}_{c}}}{{\ell}(F_{c}-G_{c})}dt, \nb\\
F_{c} &\equiv&  F\left(t + f(|y_{c}|)\right),\;\;\;
G_{c} \equiv  G\left(t - f(|y_{c}|\right), \nb\\
a_{c}(\tau) &\equiv& \frac{1}{{\ell}(F_{c}-G_{c})}, \;\;\;
b_{c}(\tau) \equiv \frac{1}{{\ell}(F_{c}+ G_{c})}, 
\eqn
we find 
\bqn
\lb{3.14}
\frac{\dot{b}_{c}}{b_{c}^{2}} &=& 
\frac{\epsilon_{c}}{a_{c}}\left(\ell^{2} + H^{2}_{c}\right)^{1/2},\nb\\
H_{c} &\equiv& \frac{\dot{a}_{c}}{a_{c}}, \;\;\; \epsilon_{c} = \pm 1.
\eqn
Eq.(\ref{2.23}) can be written as
\bqn
\lb{3.15a}
\left(H_{c}^{2} + \ell^{2}\right)^{1/2} &=& \frac{1}{6}\kappa_{5}
\epsilon_{c}\epsilon^{c}_{f}\left({\rho}_{c}  + \bar{\lambda}_{c}\right),\\
\lb{3.15b}
\frac{\dot{H}_{c}}{\left(H_{c}^{2} + \ell^{2}\right)^{1/2}} 
&=& -  \frac{1}{2} \kappa_{5}
\epsilon_{c}\epsilon^{c}_{f}\left({\rho}_{c} 
+ {p}_{c}\right),
\eqn
where $\dot{H}_{c} \equiv dH_{c}/d\tau,\; \epsilon^{c}_{f} 
= f'(y_{c})/|f'(y_{c})|$, 
$\bar{\lambda}_{c}  \equiv   \lambda_{c}/\kappa_{5}$, and 
\bq
\lb{3.16}
\epsilon_{c} =\cases{ -1, &  ${F'}_{c} > 0,\; {G'}_{c} > 0$,\cr
+ 1, &  ${F'}_{c} < 0,\; {G'}_{c} < 0$.\cr}
\eq
Note that, in order to have $\rho_{c} \ge 0$ for any given 
$\bar{\lambda}_{c}$, we must choose $\epsilon_{c}$
and $\epsilon^{c}_{f}$ such that
\bq
\lb{3.17}
\epsilon_{c}\epsilon^{c}_{f} = 1,
\eq
a choice that we shall assume in the following discussions.  
To  study the above system of differential equations further, let us 
consider the case where
\bq
\lb{3.18}
{\lambda}_{c} = 0, \;\;\;
p_{c} = w {\rho}_{c}, 
\eq
with   $w$ being an arbitrary constant. 

\subsubsection{$ w > -1 $}

In this case,  Eqs.(\ref{3.15a}) and (\ref{3.15b}) have the 
  general solution,
\bqn
\lb{3.18a}
a_{c}(\tau) &=& C_{0} 
 \sin\left[3{\ell}(1+w)\tau\right]^{\frac{1}{3(1+w)}},\nb\\
\rho_{c}(\tau) &=& \frac{6 {\ell}}{\kappa_{5}
\sin\left[3{\ell}(1+w)\tau\right]},  
\eqn
where $C_{0}$ is a positive constant. From the above expressions
we can see that in this case the 
universe starts to expand from a big-bang like singularity
at $\tau = 0$, until the moment $\tau_{max} \equiv \pi/[6{\ell}(1+w)]$. 
From that moment on the universe starts to contract
until the moment  $\tau_{f} \equiv \pi/[2{\ell}(1+w)]$, where the whole 
universe is crushed into a spacetime singularity.

For $-1 < w < -2/3$ the universe experiences an accelerating period. 
In fact, from Eq.(\ref{3.18a}) we find that
\bq
\lb{3.19a}
\ddot{a}_{c}(\tau) = - C_{0}{\ell}^{2}\frac{(2 + 3w) 
+ \sin^{2}\left[3{\ell}(1+w)\tau\right]}
{\sin^{\frac{6w+5}{3(1+w)}}\left[3{\ell}(1+w)\tau\right]},
\eq
from which we can see that the universe is accelerating during 
the time $ \tau_{0} < \tau < \pi - \tau_{0}$
for $w < -2/3$, where
\bq
\lb{3.20}
\tau_{0}  \equiv \frac{1}{3{\ell}(1- |w|)}{\mbox{Arcsin}}
\left(\sqrt{3|w| - 2}\right).
\eq
 
To study the global properties of the solutions, we first
notice that 
\bqn
\lb{3.20a}
t_{c}(t) &=& F_{c}(t) + G_{c}(t) = \frac{1}{\ell b_{c}},\nb\\
z_{c}(t) &=& F_{c}(t) - G_{c}(t)  = \frac{1}{\ell a_{c}}\nb\\
&=& \frac{1}{\ell C_{0} }\sin\left[{\ell}(1+w)\tau\right]^{-1/[3(1+w)]},
\eqn
as we can see from Eqs.(\ref{3.5a}) and (\ref{3.13}). Then,
we find
\bq
\lb{3.20b}
\frac{dz_{c}}{dt_{c}} = \epsilon_{c}\cos\left[3(1+w)\ell\tau\right].
\eq
From Eqs.(\ref{3.20a}) and (\ref{3.20b}) we obtain 
\bq
\lb{3.20c}
\epsilon_{c} t_{c} = \cases{ \infty, & $\tau = 0$,\cr
0, & $\tau = \tau_{max}$,\cr 
- \infty, & $\tau = \tau_{f}$.\cr}
\eq
In Fig. 6 the vertical line $GPH$ represents the visible 3-brane located at
$y = y_{c}$. For $ \epsilon_{c} = -1$, the 3-brane starts to expand from
the initial moment $\tau = 0$, denoted by the point $G$ in the figure,
which is singular and corresponds to ($t_{c}, z_{c})_{G} = (-\infty, \infty)$
in the ($t,z$)-plane [cf. Fig. 4]. The
expansion is continuous until the moment $\tau = \tau_{max}$, where
($t_{c}, z_{c})_{P} = (0, z_{c}^{min}$), with $z_{c}^{min} \equiv
1/(\ell C_{0})$. Afterward, the universe starts to collapse and
 a spacetime singularity of the 3-brane is finally
formed  at the moment $\tau_{f}$, where   ($t_{c}, z_{c})_{H} = 
(\infty, \infty$).  The proper time of this whole process is finite and
given by $\tau_{f}$. It should be noted that, unlike the RS1 solution,
the extension beyond the  two points $G$ and $H$ are not needed
because now they are spacetime singularities of the 3-brane. It is also
interesting to note that the five-dimensional bulk is not singular at all
even at the points $G$ and $H$.

For $ \epsilon_{c} = +1$, the evolution of the 
visible 3-brane follows the same curve $GPH$, but now the whole process
is inverse in the ($t, z$)-plane. That is, the visible 3-brane starts to 
expand from the singular point $H$ until its maximal point $P$, and then
collapses into a spacetime singularity at the point $G$.  

%%%%%%%%%%%%%%%%%%%%%%%%%%%%%%%%%%%%%%%%%%%%%%%%%%%%%%%%%%%%%%%%%%%%%%%%%%%%%%%
\begin{figure}
\includegraphics[width=\columnwidth]{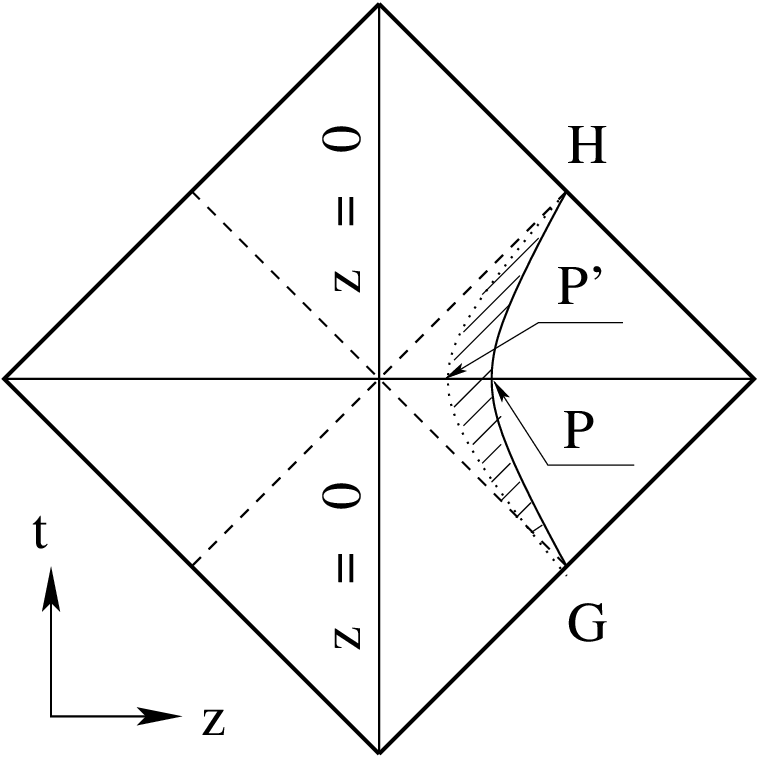}
\caption{The Penrose diagram for the case $w > -1$ of the solutions (\ref{3.18a})
in the ($t, z$)-plane of the horospheric coordinates, defined by Eq.(\ref{eq3.4}). 
The evolution of the visible 3-brane, located on the hypersurface $y = y_{c}$, is 
denoted by the vertical curve $GPH$, and the evolution of the invisible 3-brane,
located at $y = 0$, is given by the dotted curve $GP'H$. The five-dimensional bulk 
is the region between the two vertical curves. The spacetimes of the 3-branes are 
singular at the points $G$ and $H$, but not the bulk.}
\label{fig6}
\end{figure} 
%%%%%%%%%%%%%%%%%%%%%%%%%%%%%%%%%%%%%%%%%%%%%%%%%%%%%%%%%%%%%%%%%%%%%%%%%%%%%%%
 
Repeating the above analysis for the invisible 3-brane at $y = 0$, it is not
difficult to see that its trajectory in the ($t, \; z$)-plane of the horospherical
coordinates is the dotted curve $GP'H$ in Fig. 6. So, the whole bulk is the region
between the two curves $GP'H$ and $GPH$.

\subsubsection{$ w = -1 $}

In this case, we have
\bqn
\lb{3.18b}
a_{c}(\tau) &=& C_{0} e^{H_{c}\tau},\nb\\
\rho_{c}(\tau) &=& - p_{c} = \frac{6}{\kappa_{5}} \left(H_{c}^{2} 
+ {\ell}^{2}\right)^{1/2},  
\eqn
where now $H_{c}$ is a constant. Clearly, in the present case the universe is a de 
Sitter space, which is  expanding exponentially. Inserting Eq.(\ref{3.18b}) into
Eq.(\ref{3.14}) and then integrating the resulting expression, we find
\bq
\lb{3.18ba}
b_{c}(\tau) = \epsilon_{c}\frac{C_{0}H_{c}}{\left(H^{2}_{c} 
+ \ell^{2}\right)^{1/2}}e^{H_{c}\tau}.
\eq
Thus, we obtain
\bqn
\lb{3.18bb}
\frac{z_{c}}{t_{c}} &=& \epsilon_{c}\frac{H_{c}}{\left(H^{2}_{c} 
+ \ell^{2}\right)^{1/2}},\nb\\
z_{c} &=& \frac{1}{\ell C_{0}}e^{-H_{c}\tau} \nb\\
&=& \cases{ \infty, & $\tau - \infty$,\cr
0, & $\tau + \infty$,\cr}\nb\\
\epsilon_{c} t_{c} &=& \frac{\left(H^{2}_{c} + \ell^{2}\right)^{1/2}}{C_{0}H_{c}}
 e^{-H_{c}\tau} \nb\\
&=& \cases{ \infty, & $\tau - \infty$,\cr
0, & $\tau + \infty$.\cr}
\eqn
Fig. 7 shows the trajectory of the visible 3-brane in the plane of the horospheric 
coordinates $t$ and $z$   for $\epsilon_{c} = -1$. The 3-brane
starts from the moment $\tau = -\infty$ or $(t_{c}, z_{c}) = (-\infty, \infty)$, 
denoted by  the point $B$ in the figure, and ends at $\tau = \infty$, where 
$(t_{c}, z_{c}) = (0, 0)$, as shown by the vertical solid curve $BO$. 

%%%%%%%%%%%%%%%%%%%%%%%%%%%%%%%%%%%%%%%%%%%%%%%%%%%%%%%%%%%%%%%%%%%%%%%%%%%%%%%
\begin{figure}
\includegraphics[width=\columnwidth]{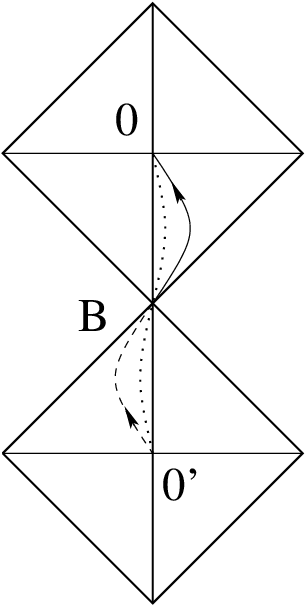}
\caption{The Penrose diagram for the case $w = -1$ and $\epsilon_{c} = -1$
of the solution (\ref{3.18b}) in the ($t, z$)-plane of the 
horospheric coordinates, defined by Eq.(\ref{eq3.4}). The evolution of the
visible 3-brane, located on the hypersurface $y = y_{c}$, is denoted by the vertical 
curve $O'BO$, and the trajectory of the invisible brane, located
at $y = 0$, is given by the dotted curve $O'BO$. The five-dimensional bulk is the 
region between the two vertical curves. From the point of view of observers who 
live on the brane, these trajectories are geodesically complete and no extension 
is needed. But, from the point of view of ``observers" in the five-dimensional bulk, 
the five-dimensional spacetime is not geodesically complete, and extension is needed. 
A possible extension is given by Fig. 8. }
\label{fig7}
\end{figure} 
%%%%%%%%%%%%%%%%%%%%%%%%%%%%%%%%%%%%%%%%%%%%%%%%%%%%%%%%%%%%%%%%%%%%%%%%%%%%%%%

It should be noted that the spacetime described by the metric
\bq
\lb{3.18bc}
ds^{2}_{4}  = d\tau^{2} - C^{2}_{0}e^{2H_{c}\tau} d\Sigma_{0}^{2},
\eq
is not geodesically complete. As a matter of fact, it covers only half of the 
hyperboloid \cite{HE73},
\bq
\lb{3.18bd}
- V^{2} + W^{2} + X^{2} + Y^{2} + Z^{2} = H^{-2}_{c},
\eq
with 
\bq
\lb{3.18be}
V + W = H^{-1}_{c}e^{H_{c}\tau} > 0.
\eq
To extend the spacetime to the other half of the hyperboloid, we introduce
the coordinates $\hat{\tau}$ and $\chi$ via the relations,
\bqn
\lb{3.18bf}
\tau &=& H^{-1}_{c}\ln\left[H_{c}\left(V+W\right)\right],\nb\\
r &=& \frac{1}{C_{0}H^{2}_{c}(V+W)} 
\cosh\left(H_{c}\hat{\tau}\right)\sin\chi,
\eqn 
where
\bq
\lb{3.18bg}
V + W = \frac{1}{H_{c}} \left[\sinh\left(H_{c}\hat{\tau}\right) 
         + \cosh\left(H_{c}\hat{\tau}\right)\cos\chi\right].
\eq
Then, we find that the metric (\ref{3.18bc}) becomes
\bq
\lb{3.18gh}
ds^{2}_{4}  =  d\hat{\tau}^{2} 
- H^{-2}_{c}\cosh^{2}\left(H_{c}\hat{\tau}\right)
\left(d\chi^{2} + \sin^{2}\chi d\Sigma_{0}^{2}\right),
\eq
which covers the whole hyperboloid  for $- \infty < \hat{\tau} < \infty$, 
$\; 0 \le \chi, \; \theta \le \pi$, and $0 \le \phi \le 2\pi$. 

On the other hand, from Eqs.(\ref{3.18bb}) and (\ref{3.18bf}) we find 
\bqn
\lb{3.18gi}
z_{c} &=& \frac{1}{\ell C_{0}H_{c}\left(V+W\right)}, \nb\\
t_{c} &=& \frac{\epsilon_{c} \left(H^{2}_{c} + \ell^{2}\right)^{1/2}}
{\ell C_{0}H_{c}^{2}\left(V+W\right)}.
\eqn
Thus, across the surface $V + W = 0$, both $z_{c}$ and $t_{c}$ change
their signs. Then, the extended region must be the one given by the vertical
dashed curve $O'B$ in Fig. 7. Therefore, in the present case the geodesically
complete spacetime of the visible 3-brane, described by metric (\ref{3.18gh}),
is that of the segment $O'BO$. 

However, as shown at beginning of this section, the spacetime
described by  metric (\ref{eq3.5}) is not geodesically complete, and one
possible extension is that given by Fig. 5. Clearly, in the present case, 
the five-dimensional bulk described by the solutions of Eqs.(\ref{3.6}) and 
(\ref{3.18b}) is also not geodesically complete. A possible extension is  
given by  Fig. 8. It is interesting to note that the segment $O'BO$ is geodesically 
complete from the point of view of observers who live on the visible 3-brane, but
this is no longer true from the point of view of the five-dimensional bulk. For 
the latter, the  geodesically complete bulk is that given by Fig. 8, in which we
have infinite number of geodesically complete visible 3-branes. 

%%%%%%%%%%%%%%%%%%%%%%%%%%%%%%%%%%%%%%%%%%%%%%%%%%%%%%%%%%%%%%%%%%%%%%%%%%%%%%%
\begin{figure}
\includegraphics[width=\columnwidth]{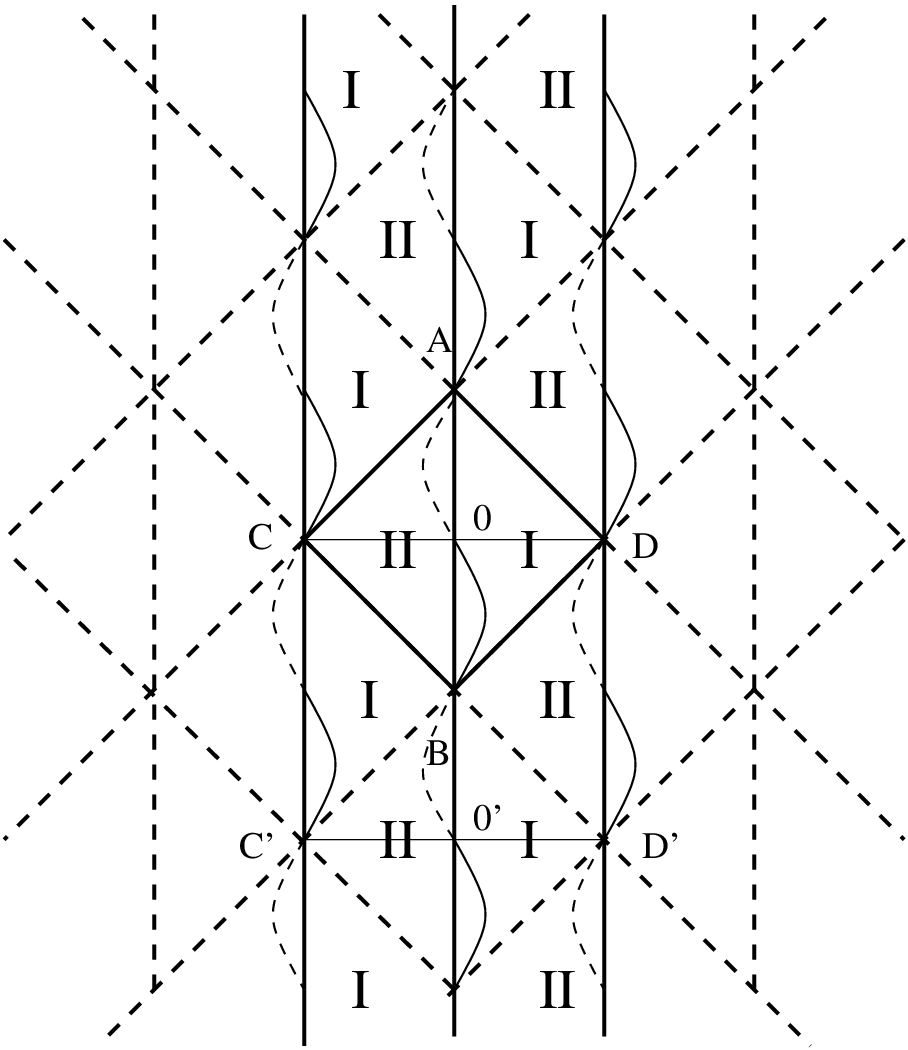}
\caption{The Penrose diagram of  the extended spacetime for $w = -1$
and $\epsilon_{c} = -1$, given by
Eq.(\ref{3.18b}). The vertical curves,  both solid and dashed,  represent the 
history of a visible 3-brane. Each segment, consisting of a solid and a dashed
part, is geodesically complete for observers living on the visible brane. The
trajectory of the invisible 3-brane is represented by similar vertical curves, and
the five-dimensional bulk is the region between the two curves, as shown in Fig. 7.   }  
\label{fig8} 
\end{figure} 
%%%%%%%%%%%%%%%%%%%%%%%%%%%%%%%%%%%%%%%%%%%%%%%%%%%%%%%%%%%%%%%%%%%%%%%%%%%%%%%

When   $\epsilon_{c} = + 1$, the global properties of the corresponding solution
can be obtained from the case of $\epsilon_{c} = -1$ by the replacement $t_{c}
\rightarrow - t_{c}$. 

Similarly, it can be shown that the trajectory of the invisible brane located
at $y = 0$ is similar to that of the visible brane, and is given by the dotted
curve $O'BO$ in Fig. 7. The five-dimensional bulk is the region between the two 
vertical curves.

\subsubsection{$ w < -1 $}

In this case, we have
\bqn
\lb{3.18c}
a_{c}(\tau) &=& \frac{C_{0}}{\sin\left[3{\ell}\left(\left|w\right| 
- 1\right)\tau\right]^{\frac{1}{3(|w| - 1)}}},\nb\\
\rho_{c}(\tau) &=& \frac{6 {\ell}}{\kappa_{5}
\sin\left[3{\ell}(|w|-1)\tau\right]},
\eqn
from which we can see that the universe is singular at $\tau = 0$, 
where the radius of the universe is infinitely large $a_{c}(0) = \infty$.
As time increases, the radius of the universe decreases until the moment 
$\tau = \tau_{min}$ when it reaches its minimum $a_{c}(\tau_{min}) = 
a_{c}^{0}$, where
\bq
\lb{3.21}
\tau_{min}  \equiv \frac{\pi}{6{\ell}(|w| - 1)}.
\eq
After this moment, the radius of the universe starts to increase until 
the moment $\tau = 2\tau_{min}$, at
which it becomes infinitely large, and the spacetime becomes singular. 
Note that in this case the acceleration 
is always positive,  
\bq
\lb{3.19b}
\ddot{a}_{c}(\tau) =  C_{0}{\ell}^{2}\frac{3(|w|-1) 
+ \cos^{2}\left[3{\ell}(|w|-1)\tau\right]}
{\sin^{\frac{6|w| - 5}{3(|w| - 1)}}\left[3{\ell}(|w|-1)\tau\right]}.
\eq

On the other hand, we have
\bqn
\lb{3.20aa}
\frac{dz_{c}}{dt_{c}} &=& - \epsilon_{c}\cos\left[3(|w| -1)\ell\tau\right],\nb\\
z_{c}(t) &=& \frac{1}{\ell C_{0} }\sin\left[3{\ell}(|w|-1)
\tau\right]^{1/[3(|w|-1)]}\nb\\
&=& \cases{0, & $\tau = 0$,\cr
(\ell C^{0})^{-1}, & $\tau = \tau_{min}$,\cr
0, & $\tau = 2\tau_{min}$.\cr}
\eqn
Thus, in the $(t_{c}, z_{c})$-plane of the horospherical coordinates,
for the case $\epsilon_{c} = -1$ the visible 3-brane starts to collapse
from the singular point $(t_{c}, z_{c}) = (0, 0)$ until the point 
$z_{c}(\tau_{min}) = (\ell C_{0})^{-1}$.
Afterward, it starts to expand upto the moment $\tau = 2\tau_{in}$,
where $(t_{c}, z_{c}) = (t_{c}^{0}, 0)$, at which a spacetime singularity
is developed [cf. Fig. 9]. Since the 3-brane is singular at the points
$O$ and $P$, the spacetime of the visible 3-brane is not extendible beyond
these two points.

The global properties for the case of   $ \epsilon_{c} = +1$ can be obtained
from the one of $ \epsilon_{c} = -1$ by the replacement $t_{c} \rightarrow
- t_{c}$.

Similarly, it can be shown that the trajectory of the invisible 3-brane is denoted 
by the dotted curve $OP$ in Fig. 9.  

%%%%%%%%%%%%%%%%%%%%%%%%%%%%%%%%%%%%%%%%%%%%%%%%%%%%%%%%%%%%%%%%%%%%%%%%%%%%%%%
\begin{figure}
\includegraphics[width=\columnwidth]{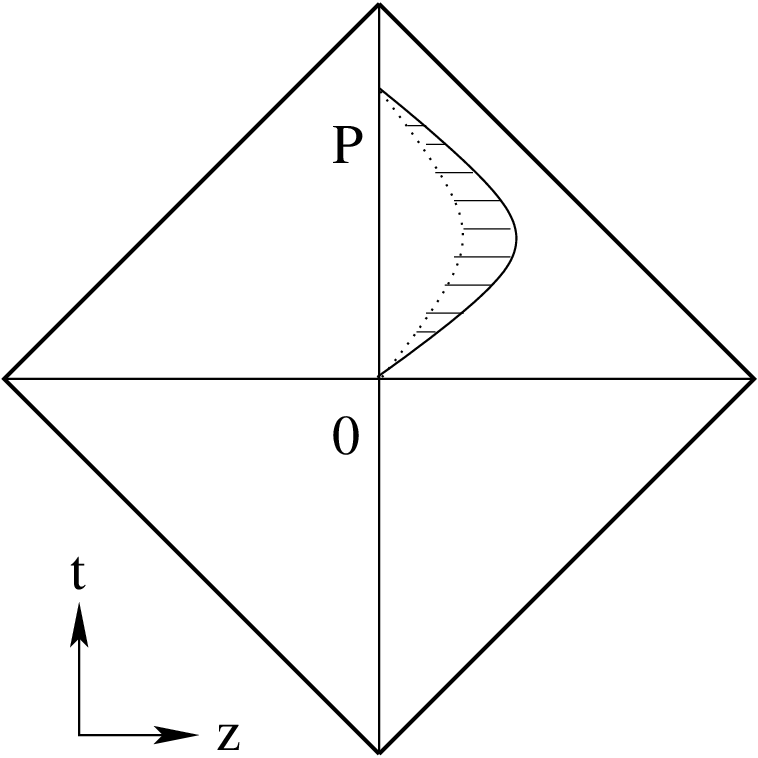}
\caption{The Penrose diagram of  the extended spacetime for $w < -1$ and
$\epsilon_{c} = -1$ of the solutions (\ref{3.18c}) in the ($t, z$)-plane of the 
horospheric coordinates. The trajectory of the  visible (invisible) 3-brane
is denoted by the solid (dotted) curve $OP$. The spacetimes of the visible and
invisible 3-branes are singular at the points $O$ and $P$.   }  
\label{fig9} 
\end{figure} 
%%%%%%%%%%%%%%%%%%%%%%%%%%%%%%%%%%%%%%%%%%%%%%%%%%%%%%%%%%%%%%%%%%%%%%%%%%%%%%%

\section{Two 3-Branes in vacuum Bulk   with $k = 0$}

\renewcommand{\theequation}{4.\arabic{equation}}
\setcounter{equation}{0} 

When the bulk is vacuum and the spatial sectors of the two 
3-branes are flat, $k = 0$, the corresponding solutions of the
five-dimensional bulk are well-known \cite{Taub51,BCG00}, and
they can be divided into three different classes, which will be
referred to, respectively, as Class IA, Class IB, and Class II
solutions. In the following we consider them separately.

\subsection{Class IA Solutions}

This class of solutions, after the replacement of Eq.(\ref{2.8}),
is given by 
\bqn
\lb{4.1a}
N^{2}(t, |y|) &=& \frac{F G' }{G^{2/3}}, \nb\\
A^{2}(t, |y|) &=&  G^{2/3}, \nb\\
B^{2}(t, |y|) &=& \frac{F G' }{G^{2/3}}{f'(|y|)}^{2},
\eqn
where $F $ and $G$ are given by Eq.(\ref{3.6a}), with  $f(|y|)$ 
being arbitrary function of $|y|$ only. As first noticed by Taub 
\cite{Taub51}, the spacetime outside the
3-branes in this case is flat. In fact, introducing the  
coordinates $T$ and $ X^{i}\; (i = 1, 2, 3, 4)$ via the relations,
\bqn
\lb{eq4.1}
T - X^{1} &=& G^{1/3}\left(t - f(y)\right),\nb\\
T + X^{1} &=& 3\int^{t + f(y)}{F(x) dx} \nb\\
& &
+ G^{1/3}\left(t - f(y)\right)\left[\left(x^{2}\right)^{2}  
 + \left(x^{3}\right)^{2} + \left(x^{4}\right)^{2}\right],\nb\\
X^{i} &=& G^{1/3}\left(t - f(y)\right) x^{i}, \; (i = 2, 3, 4),
\eqn
the metric outside the 3-branes will take the Minkowski form, 
$ ds^{2}_{5} = \eta_{AB}dX^{A}dX^{B}$. From the above expressions we
find that
\bq
\lb{eq4.2}
T^{2} - R^{2}  = 3  G^{1/3}\left(t^{2} - f(y)\right)\int^{t + f(y)}{F(x) dx},
\eq
where $R^{2} \equiv \left(X^{1}\right)^{2} + \left(X^{2}\right)^{2} 
+ \left(X^{3}\right)^{2} + \left(X^{4}\right)^{2}$.

Introducing $\tau$ and $a_{c}(\tau)$ via the relations,
\bqn
\lb{4.3}
& & d\tau = \left(\frac{F_{c} {G'}_{c}}{ {G}_{c}^{2/3}}\right)^{1/2} dt,\nb\\
& & a_{c}(\tau) = G_{c}^{1/3}(t),
\eqn
we find that Eq.(\ref{2.23}) yields,
\bqn
\lb{4.4}
\frac{\dot{H}_{c}}{H_{c}} &=& \frac{1}{2}\kappa_{5}\epsilon_{c}
\left(\rho_{c} + p_{c}\right),\nb\\
 H_{c} &=& -\frac{1}{6}\kappa_{5}\epsilon_{c}
\left(\rho_{c} + \bar{\lambda}_{c}\right),
\eqn
where, as previously, $H_{c} \equiv \dot{a}_{c}/a_{c}$, 
$\dot{H}_{c} \equiv dH_{c}/d\tau$, and 
$\epsilon_{c} \equiv f'(|y_{c}|)/|f'(|y_{c}|)|$.
It is remarkable to note that $F_{c}(t)$ does not appear explicitly in 
Eq.(\ref{4.4}).

When $\lambda_{c} = 0$ and $p_{c} = w\rho_{c}$, Eq.(\ref{4.4}) has the 
solution,
\bq
\lb{4.5}
a_{c}(\tau) = \tau^{\frac{1}{3(1+w)}},
\eq
from which we find that
\bqn
\lb{4.6}
\ddot{a}_{c}(\tau) &=& - \frac{3w + 2}{9(1+w)^{2}} 
\tau^{-\frac{6w + 5}{3(1+w)}},\nb\\
\rho_{c}(\tau) &=& - \frac{2\epsilon_{c}}{\kappa_{5}(1+w) \tau}.
\eqn
Clearly, to have $\rho_{c}$ non-negative, we much choose $\epsilon_{c}$
such that
\bq
\lb{4.7}
\epsilon_{c} = \cases{ - 1, & $ w > -1$,\cr
+ 1, & $ w < -1$.\cr}
\eq
From Eq.(\ref{4.6}) we can also see that the universe is accelerating for
$w < -2/3$. 

It should be noted that Eqs.(\ref{4.5}) and (\ref{4.6}) hold only for 
$w \not= -1$. When $w = -1$ from Eq.(\ref{4.4}) we find that 
\bqn
\lb{4.8}
a_{c}(\tau) &=& a_{0}e^{H_{c} \tau},\nb\\
\rho_{c}(\tau) &=& - p_{c}(\tau) = - \frac{ 6 \epsilon_{c}H_{c}}{\kappa_{5}},
\; (w  = -1), 
\eqn
where $a_{0}$ and $H_{c}$ are positive constants.    Obviously,
the spacetime of the visible 3-branes is always acceleratingly expanding. 
%%%%%%%%%%%%%%%%%%%%%%%%%%%%%%%%%%%%%%%%%%%%%%%%%%%%%%%%%%%%%%%%%%%%%%%%%%%%%%
%But, depending on the signs of $\epsilon_{c}$, the spacetime of the 3-brane 
%at $y = y_{c}$ could be either de Sitter ($\epsilon_{c} = -1$) or anti-de 
%Sitter ($\epsilon_{c} = 1$).
%%%%%%%%%%%%%%%%%%%%%%%%%%%%%%%%%%%%%%%%%%%%%%%%%%%%%%%%%%%%%%%%%%%%%%%%%%%%%%%

On the other hand, from Eq.(\ref{eq4.2}) we find that  the
visible 3-brane in the five-dimensional Minkowski coordinates $T$ and $X^{i}$
is a hyperboloid,
\bqn
\lb{eq4.3}
\left[T^{2}  - R^{2} \right]_{y=y_{c}}  &=&
3  G^{1/3}\left(t - f(y_{c})\right)\nb\\
& & \times \int^{t + f(y_{c})}{F(x) dx}.
\eqn
Similarly, the invisible 3-brane located at $y = 0$ is also a hyperboloid,
\bqn
\lb{eq4.4}
\left[T^{2}  - R^{2} \right]_{y=0}  &=&
3  G^{1/3}\left(t - f(0)\right)\nb\\
& & \times \int^{t + f(0)}{F(x) dx}.
\eqn
The two 3-branes will collide at the moment $t_{d}$, where $t_{d}$ is a
solution of the equation,
\bq
\lb{eq4.5}
\left[\frac{G\left(t - f(y_{c})\right)}{G\left(t - f(0)\right)}\right]^{1/3}
= \frac{\int^{t + f(y_{c})}{F(x) dx}}{\int^{t + f(0)}{F(x) dx}}.
\eq
The bulk is the region between these two hyperboloids. However, since it
is flat, no radiation,  neither gravitational nor non-gravitational, from the
two 3-branes is leaked into the bulk, although there may be energy exchange between 
the two 3-branes at the moment when they collide [cf. Fig. 10].

%%%%%%%%%%%%%%%%%%%%%%%%%%%%%%%%%%%%%%%%%%%%%%%%%%%%%%%%%%%%%%%%%%%%%%%%%%%%%%%
\begin{figure}
\includegraphics[width=\columnwidth]{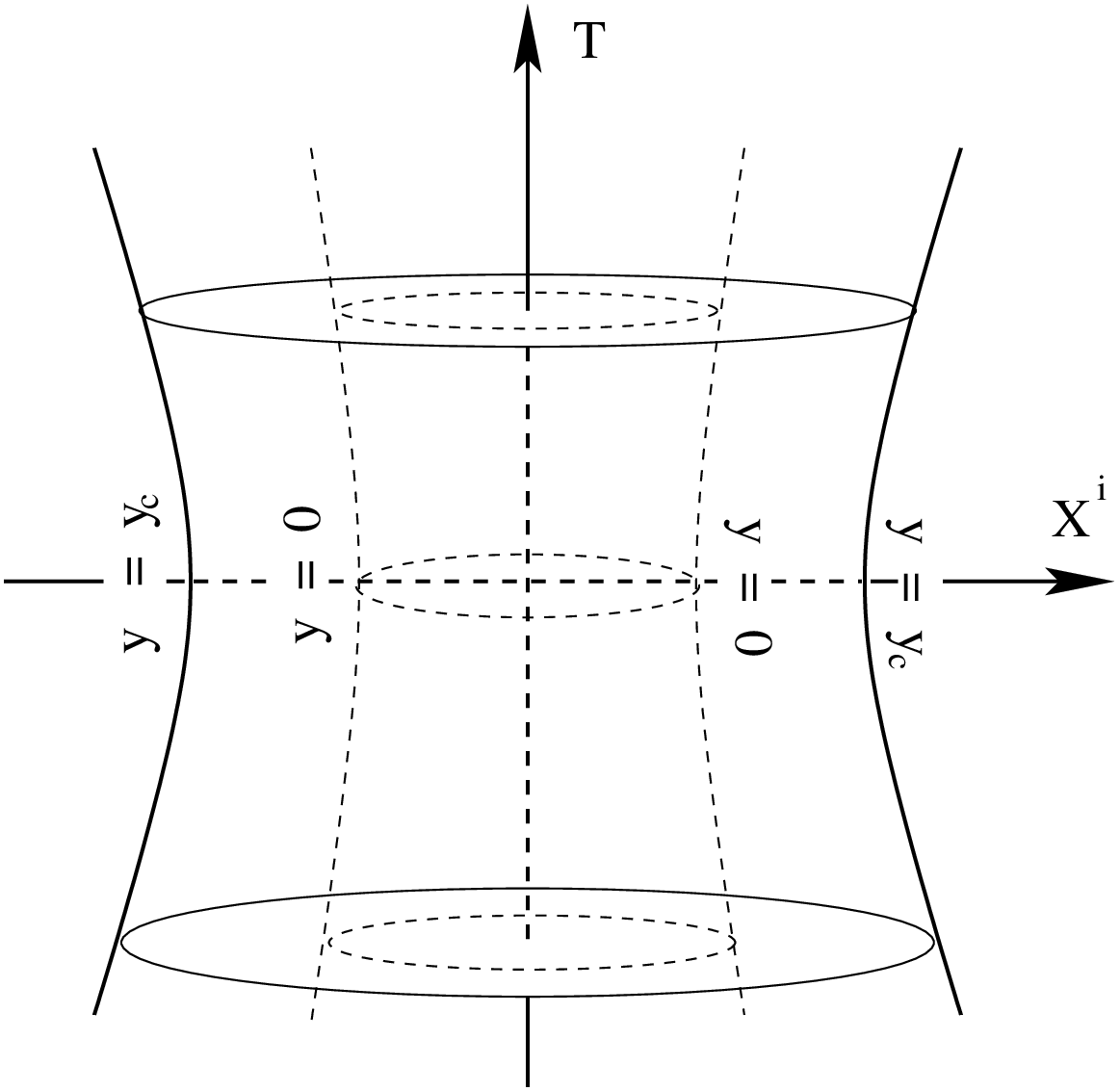}
\caption{The two 3-branes in the Minkowski coordinates $T$ and $X^{i} \;
(i = 1, 2, 3, 4)$, defined by Eq.(\ref{eq4.1}), at a given time $t$. 
Each of the two branes is a hyperboloid. In some cases the two branes 
collide, provided that Eq.(\ref{eq4.5}) has real solutions.      }  
\label{fig10} 
\end{figure} 
%%%%%%%%%%%%%%%%%%%%%%%%%%%%%%%%%%%%%%%%%%%%%%%%%%%%%%%%%%%%%%%%%%%%%%%%%%%%%%%

\subsection{Class IB Solutions}

In this case, the solutions are given by
\bqn
\lb{4.1b}
N^{2}(t, |y|) &=& \frac{F' G}{F^{2/3}}, \nb\\
A^{2}(t, |y|) &=&  F^{2/3}, \nb\\
B^{2}(t, |y|) &=& \frac{F' G}{F^{2/3}}{f'(|y|)}^{2}.
\eqn
Introducing $\tau$ and $a_{c}(\tau)$ via the relations,
\bqn
\lb{4.9}
& & d\tau = \left(\frac{{F'}_{c} G_{c}}{ {F}_{c}^{2/3}}\right)^{1/2} dt,\nb\\
& & a_{c}(\tau) = F_{c}^{1/3}(t),
\eqn
we find that Eq.(\ref{2.23}) yields,
\bqn
\lb{4.10}
\frac{\dot{H}_{c}}{H_{c}} &=& - \frac{1}{2}\kappa_{5}\epsilon_{c}
\left(\rho_{c} + p_{c}\right),\nb\\
 H_{c} &=& \frac{1}{6}\kappa_{5}\epsilon_{c}
\left(\rho_{c} + \bar{\lambda}_{c}\right).
\eqn
Comparing Eq.(\ref{4.10}) with Eq.(\ref{4.4}) we can see that we can obtain one   
from the other by simply replacing $\epsilon_{c}$ by $- \epsilon_{c}$. Thus,
the physics of the solutions in the present case can be deduced from the ones given
in the last case. 
Therefore, in the following we shall not consider this case any further.

\subsection{Class II Solutions}

In this case, the corresponding solutions are given by
\bqn
\lb{4.2}
N^{2}(t, |y|) &=& \frac{F' G' }{\left(F + G\right)^{2/3}}, \nb\\
A^{2}(t, |y|) &=&  \left(F + G\right)^{2/3}, \nb\\
B^{2}(t, |y|) &=& \frac{F' G' }{\left(F + G\right)^{2/3}}{f'(|y|)}^{2}.
\eqn
Before studying the dynamics of the 3-branes, let us first consider the
global structure of the spacetime described by the above solutions without
3-branes, that is, the solutions of Eq.(\ref{4.2}) with $|y|$ being replaced
by $y$ where we temporarily extend the range of $y$  to $y \in (-\infty, \infty)$. 
Then, by the coordinate transformation,
\bq
\lb{4.2aa}
T \equiv F + G, \;\;\; Y \equiv F - G,
\eq
the corresponding metric takes the form,
\bq
\lb{4.2ab}
ds^{2}_{5} = \frac{1}{4T^{2/3}}\left(dT^{2} - dY^{2}\right) 
- T^{2/3}d\Sigma^{2}_{0}.
\eq
Clearly, the 5D spacetime is singular at $T = 0$, and the corresponding 
Penrose  diagram is given by Fig. 11.

%%%%%%%%%%%%%%%%%%%%%%%%%%%%%%%%%%%%%%%%%%%%%%%%%%%%%%%%%%%%%%%%%%%%%%%%%%%%%%%
\begin{figure}
\includegraphics[width=\columnwidth]{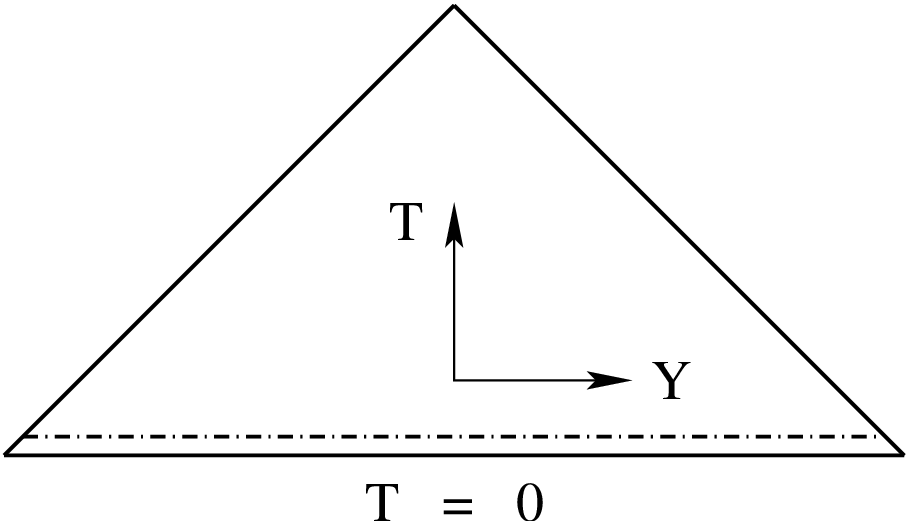}
\caption{The Penrose diagram of the solutions of Eq.(\ref{4.2ab}) 
in the $(T, Y)$-plane,
defined by Eq.(\ref{4.2aa}). The five-dimensional 
spacetime is singular at $T = 0$.}  
\label{fig11} 
\end{figure} 
%%%%%%%%%%%%%%%%%%%%%%%%%%%%%%%%%%%%%%%%%%%%%%%%%%%%%%%%%%%%%%%%%%%%%%%%%%%%%%%

Introducing $\tau$, $a_{c}(\tau)$ and $b_{c}(\tau)$ via the relations,
\bqn
\lb{4.11}
& & d\tau = \left(\frac{{F'}_{c} {G'}_{c}}
{\left( F_{c} + {G}_{c}\right)^{2/3}}\right)^{1/2} dt,\nb\\
& & a_{c}(\tau) = \left(F_{c} + {G}_{c}\right)^{1/3},\nb\\
& & b_{c}(\tau) = \left( F_{c} - {G}_{c}\right)^{1/3},
\eqn
from Eq.(\ref{2.23}) we find,
\bqn
\lb{4.12}
\frac{\dot{H}_{c} + 2\left(H^{2}_{c} - \Delta^{2}\right)}{\Delta} 
&=& - \frac{1}{2}\kappa_{5}\epsilon_{c}^{0}\epsilon_{c}
\left(\rho_{c} + p_{c}\right),\nb\\
 \Delta &=& \frac{1}{6}\kappa_{5}\epsilon_{c}^{0}\epsilon_{c}
\left(\rho_{c} + \bar{\lambda}_{c}\right),
\eqn
where $\epsilon_{c}^{0} = \pm 1$, and
\bqn
\lb{4.13}
& & \Delta \equiv \left(H^{2}_{c} - \frac{4}{9 a^{4}_{c}}\right)^{1/2},\nb\\
& & \dot{b}_{c}(\tau) = \frac{\epsilon_{c}^{0}a^{3}_{c}}{b^{2}_{c}}\Delta.
\eqn
%It is remarkable to note that the function $b_{c}(t)$ does not appear
%in the above equations. 
To study this case further, in the rest of 
this sub-section, we also restrict ourselves to the case where 
$\lambda_{c} = 0$ and $p_{c} = w \rho_{c}$. 
Then, from Eq.(\ref{4.12}) we find that
\bq
\lb{4.14}
\ddot{a}_{c} + (2+3w)\frac{\dot{a}_{c}^{2}}{a_{c}}
- \frac{4(1+3w)}{9 a^{3}_{c}} = 0.
\eq
To solve this equation, it is found convenient to distinguish the following cases.

\subsubsection{$ w > - \frac{1}{3}$}

In this case, introducing the function $X(\tau)$ via the relation
\bq
\lb{4.23}
a_{c}(\tau) = \beta X^{\frac{1}{3(1+w)}},
\eq
where $\beta$ is an arbitrary constant and different from the one introduced
in Eq.(\ref{2.6}),   Eq.(\ref{4.14})
reduces to,
\bq
\lb{4.25}
\ddot{X} - \frac{4(1+w)(1+3w)}{3\beta^{4}} X^{\frac{3w-1}{3(1+w)}} = 0.
\eq
Choosing the constant $\beta$ as
\bq
\lb{4.26}
\beta = \left|\frac{4(1+w)(1+3w)}{3}\right|^{1/4},
\eq
we find that Eq.(\ref{4.25}) finally takes the form
\bq
\lb{4.27}
\ddot{X} - X^{-\delta} = 0,
\eq
where
\bq
\lb{4.28}
\delta \equiv \frac{1-3w}{3(1+w)}.
\eq
It can be shown that Eq.(\ref{4.27}) allows the first integral,
\bq
\lb{4.29}
\dot{X} = \pm  \sqrt{\frac{X^{2n} + X_{1}}{n}},
\eq
where $X_{1}$ is an integration constant, and 
\bq
\lb{4.30}
n \equiv \frac{1-\delta}{2} = \frac{1+3w}{3(1+w)},
\eq
from which we find that for $-1/3 < w < \infty$ we always have $0 < n \le 1$.
To have an expanding universe, in the following we shall choose
the ``+" sign in Eq.(\ref{4.29}).

When $w = \frac{1}{3}$, Eq.(\ref{4.29}) can be integrated explicitly
and gives,
\bqn
\lb{4.31}
a_{c}(\tau) &=& \sqrt{\frac{4}{3}}\left\{\left(\tau + \tau_{0}\right)^{2}
- \tau^{2}_{0}\right\}^{1/4},\nb\\
\dot{a}_{c}(\tau) &=&  \frac{\tau + \tau_{0}}
{\sqrt{3} \left[\left(\tau + \tau_{0}\right)^{2}
- \tau^{2}_{0}\right]^{3/4}},\nb\\
\ddot{a}_{c}(\tau) &=& - \frac{\left(\tau + \tau_{0}\right)^{2}
+2 \tau^{2}_{0}}
{\sqrt{12} \left[\left(\tau + \tau_{0}\right)^{2}
- \tau^{2}_{0}\right]^{7/4}},\nb\\
\rho_{c}(\tau) &=&  \frac{3\epsilon^{0}_{c}\epsilon_{c}\tau_{0}}
{\kappa_{5}\left[\left(\tau + \tau_{0}\right)^{2}
- \tau^{2}_{0}\right]}, \; (w = 1/3), 
\eqn
where $\tau_{0}$ is a constant. Clearly, in this case the universe
starts to expand from the singular point $\tau = 0$, and the acceleration 
of the expansion is always negative. 

On the other hand, from Eqs.(\ref{4.2aa}), (\ref{4.11}) and (\ref{4.31})
we find that
\bqn
\lb{4.32aa}
Y_{c}(\tau) &=& \frac{3\epsilon^{0}_{c}\tau_{0}}{2}
\left(\frac{4}{3}\right)^{3/2}\int{\frac{d\tau}
{\left[(\tau + \tau_{0})^{2} - \tau^{2}_{0}\right]^{1/4}}},\nb\\
T_{c}(\tau) &=& \left(\frac{4}{3}\right)^{3/2}
\left[(\tau + \tau_{0})^{2} - \tau^{2}_{0}\right]^{3/4}.
%\frac{dY_{c}}{dT_{c}} &=& \epsilon^{0}_{c} \frac{\tau_{0}}
%{\tau + \tau_{0}}.
\eqn
Thus, we have $T_{c}(0) = Y_{c}(0) = 0$, and
\bqn
\lb{4.32ab}
T_{c}(\tau) &\sim& \tau^{3/2} \rightarrow \infty, \nb\\
Y_{c}(\tau) &\sim& \tau^{1/2} \rightarrow \infty, 
\eqn
as $\tau \rightarrow \infty$. 
Then, one can see that the trajectory of the visible 3-brane
in the $(T, Y)$-plane is given by the solid curve $OAP$ for 
$\epsilon^{0}_{c} = +1$ and   $OBP$ for $\epsilon^{0}_{c} = -1$, 
respectively, in Fig. 12. 

Similarly, it can be shown that the trajectory of the invisible 3-brane
in the $(T, Y)$-plane is given by the dotted curve $OA'P$ for 
$\epsilon^{0}_{c} = +1$ and   $OB'P$ for $\epsilon^{0}_{c} = -1$ in
Fig. 12. Then, the five-dimensional bulk  is the region
between the solid and dotted curves in each case. The spacetimes of
both visible and invisible 3-branes are singular at the initial moment
$\tau = 0$.

%%%%%%%%%%%%%%%%%%%%%%%%%%%%%%%%%%%%%%%%%%%%%%%%%%%%%%%%%%%%%%%%%%%%%%%%%%%%%%%
\begin{figure}
\includegraphics[width=\columnwidth]{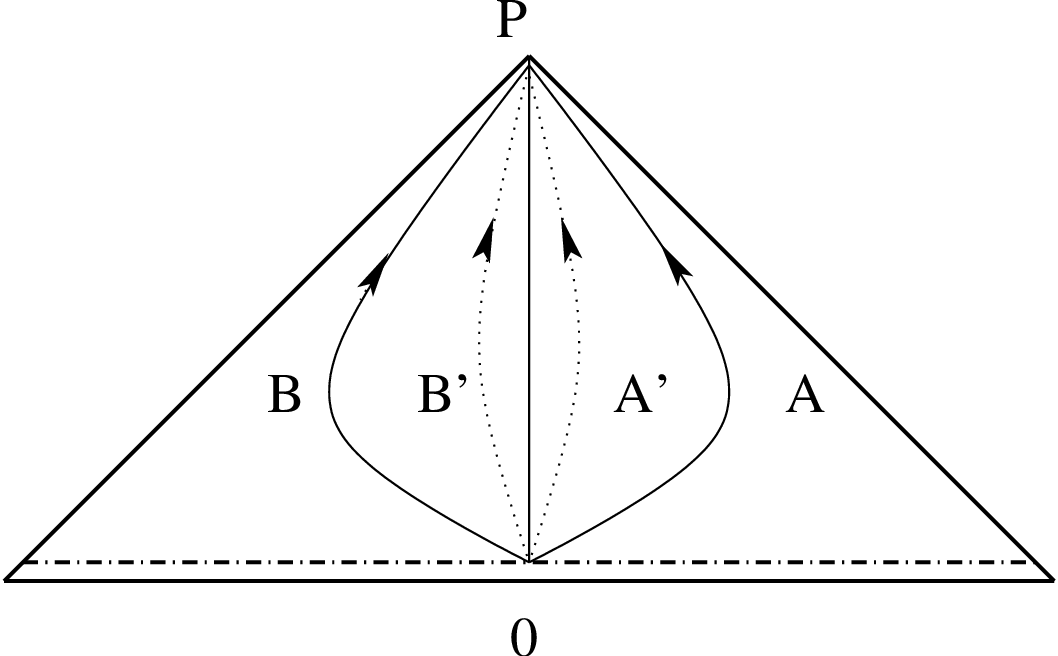}
\caption{The Penrose diagram of the Class II solutions   
in the $(T, Y)$-plane for $w > - 1$.  The solid (dotted) curves $OAP$ ($OA'P$)
and $OBP$ ($OB'P$) are the trajectories of the visible (invisible) 3-branes, 
respectively, for $\epsilon^{0}_{c} = +1$ and $\epsilon^{0}_{c} = -1$. 
The spacetimes of both visible and invisible 3-branes are singular at the  
point $O$.}  
\label{fig12} 
\end{figure} 
%%%%%%%%%%%%%%%%%%%%%%%%%%%%%%%%%%%%%%%%%%%%%%%%%%%%%%%%%%%%%%%%%%%%%%%%%%%%%%%

When $w \not= \frac{1}{3}$, using the integral,
\bqn
\lb{4.32ac}
\int{\frac{dx}{\sqrt{x^{2n} + a^{2n}}}} = \frac{x}{a^{n}}
F\left(\frac{1}{2}, \frac{1}{2n}; 1 + \frac{1}{2n}; 
- \left(\frac{x}{a}\right)^{2n}\right),\nb\\
\eqn
we find that Eq.(\ref{4.29}) has the solution,
\bq
\lb{4.32}
X \; F\left(\frac{1}{2}, \frac{1}{2n}; 1 + \frac{1}{2n}; 
- \left(\frac{X}{X_{0}}\right)^{2n}\right) = 
\left(\frac{X^{2n}_{0}}{n}\right)^{1/2}
\tau,  
\eq
where $X_{0} = X^{1/2n}_{1}$, and $F(a,b;c;z)$ denotes the ordinary
hypergeometric function with 
\bq
\lb{aa}
 F(a,b;c;0) = 1. 
\eq
Then, from Eqs.(\ref{4.23}), (\ref{4.32}) and (\ref{aa}), we find
\bq
\lb{4.32a}
a_{c}(\tau = 0) = 0.
\eq
On the other hand, using the relation,
\bqn
\lb{4.32b}
F(a,b;c;z) &=& \frac{\Gamma(c)\Gamma(b-a)}{\Gamma(b)\Gamma(c-a)}
(-z)^{-a} \nb\\
& & \times F\left(a,1-c + a;1-b+a;\frac{1}{z}\right)\nb\\
& & + \frac{\Gamma(c)\Gamma(a-b)}{\Gamma(a)\Gamma(c-b)}
(-z)^{-b} \nb\\
& & \times F\left(b,1-c + b;1-a+b;\frac{1}{z}\right),\nb\\
\eqn
we find 
\bqn
\lb{4.32c}
X \; F\left(\frac{1}{2}, \frac{1}{2n}; 1 + \frac{1}{2n}; 
- \left(\frac{X}{X_{0}}\right)^{2n}\right) &\simeq&
\frac{X^{n}_{0}}{1- n}X^{1- n} \nb\\
& & \rightarrow \infty,
\eqn
as $ X \rightarrow \infty$. Therefore, we must have
\bq
\lb{4.32d}
a_{c}(\tau = \infty) = \infty.
\eq
From the above expressions, we also find that
\bqn
\lb{4.32e}
\ddot{a}_{c}(\tau) &=& - \frac{\beta}{3(1+w)(1+3w)}  
\left(X^{- \frac{1}{1+w}} \right.\nb\\
& & \left. + (3w+2)X^{2n}_{0}X^{-\frac{6w + 5}{3(1+w)}}\right).
\eqn
Thus, in the present case the universe is always decelerating.
It can also be shown that
\bqn  
\lb{4.32g}
& & \rho_{c}(\tau) = \frac{\rho_{0}}{X},\nb\\
& & \rho_{0} \equiv \frac{6\epsilon^{0}_{c}\epsilon_{c}X^{n}_{0}}
{\sqrt{3(1+w)(1+3w)\kappa^{2}_{5}}}.
\eqn

From Eqs.(\ref{4.2aa}), (\ref{4.11}) and (\ref{4.23}), on the other
hand, we find that 
\bqn
\lb{4.32ga}
Y_{c}(\tau) &=& \epsilon_{c}^{0} \beta^{3}  X^{1/(1+w)}\nb\\
& & \times
F\left(\frac{1}{2}, \frac{3}{2(1+3w)};  \frac{5+6w}{2(1+3w)}; 
- \left(\frac{X}{X_{0}}\right)^{2n}\right),\nb\\
T_{c}(\tau) &=& a^{3}_{c}(\tau) =  \beta^{3}X^{1/(1+w)}. 
\eqn
Thus, we have $T_{c}(0) = Y_{c}(0) = 0$, and
\bqn
\lb{4.32aab}
T_{c}(\tau) &\sim& X^{1/(1+w)} \rightarrow \infty, \nb\\
Y_{c}(\tau) &\sim& \epsilon_{c}^{0} \beta^{3}
\left(\frac{3X^{n}_{0}}{2-3w}X^{\frac{2-3w}{3(1+w)}} + C_{0}\right),
\eqn
as  $\tau \rightarrow \infty$ (or $X \rightarrow \infty$),
where
\bq
\lb{4.32aac}
C_{0} \equiv \frac{3X^{1/(1+w)}_{0}}{2(1+3w)\pi^{1/2}}
\Gamma\left(\frac{3}{2(1+3w)}\right) 
\Gamma\left(\frac{3w-2}{2(1+3w)}\right).
\eq
Since $1/(1+w) > (2-3w)/[3(1+w)]$ for $w > -1/3$, we find that 
the trajectory of the visible 3-brane in the $(T, Y)$-plane is also given
by the curves $OAP$ and $OBP$ in Fig. 12, respectively, for 
$\epsilon_{c}^{0}=  1$ and  $\epsilon_{c}^{0}= - 1$. The trajectory of 
the invisible 3-brane at $y =0$ is similar to the visible 3-brane, as 
indicated by the dotted vertical curves.

\subsubsection{$w = - \frac{1}{3}$}

In this case, Eq.(\ref{4.14}) has the solution,
\bq
\lb{4.15}
a_{c}(\tau) = a_{0} \tau^{1/2},\; (w = -1/3),
\eq
where $a_{0}$ is a positive constant. Then, the corresponding energy density is
given by
\bqn
\lb{4.16}
& & \rho_{c}(\tau) = \frac{\rho^{0}_{c}}{\tau},\nb\\
& & \rho_{c}^{0} \equiv \frac{3\epsilon^{0}_{c}\epsilon_{c}}{\kappa_{5}}
\left(1 - \frac{16}{9a_{0}^{4}}\right)^{1/2}.
\eqn
It is interesting to note that choosing $a_{0} = \sqrt{3/4}$ we have 
$\rho_{c} = 0$, and the spacetime on the 3-brane becomes vacuum.  

To study the global structure of the 3-branes in this case, we first notice
that
\bqn
\lb{4.16aa}
& & T_{c}(\tau) = a^{3}_{0}\tau^{3/2},\nb\\
& & Y_{c}(\tau) = Y^{0}_{c}\tau^{3/2},\nb\\
& & Y_{c}^{0}  \equiv \epsilon^{0}_{c}a^{3}_{0} 
\left(1 - \frac{16}{9a_{0}^{4}}\right)^{1/2}.
\eqn
Since $\left|Y_{c}^{0}\right| < a^{3}_{0}$, it can be shown that the corresponding
Penrose diagram for the visible and invisible 3-branes is also given by Fig. 12.

\subsubsection{$ -1 < w < - \frac{1}{3}$} 

In this case, choosing the
constant $\beta$ as that given by Eq.(\ref{4.26}), we find that 
Eq.(\ref{4.25}) becomes 
\bq
\lb{4.33}
\ddot{X} + X^{-\delta} = 0,
\eq
where $\delta$ is still given by Eq.(\ref{4.28}). It can be shown that 
Eq.(\ref{4.33}) has the first integral,
\bq
\lb{4.34}
\dot{X} = \pm  \sqrt{\frac{X^{-2m} + X_{1}}{m}},
\eq
where    
\bq
\lb{4.35}
m \equiv \frac{\delta -1 }{2} = - \frac{1+3w}{3(1+w)} > 0.
\eq

When $w = -\frac{2}{3}$, Eq.(\ref{4.34}) has the solution,
\bq
\lb{4.36}
a_{c}(\tau) = a_{0}\left[\tau\left(\tau + \frac{4}{3a^{2}_{0}}\right)
\right]^{1/2}, \; (w = -2/3),
\eq
where $a_{0}$ is a positive constant. Then, we find that
\bqn
\lb{4.37}
\dot{a}_{c}(\tau) &=& a_{0}\frac{\tau + \frac{2}{3a^{2}_{0}}}
{\left[\tau\left(\tau + \frac{4}{3a^{2}_{0}}\right)
\right]^{1/2}},\nb\\
\ddot{a}_{c}(\tau) &=& -  \frac{4}
{9a^{3}_{0}\left[\tau\left(\tau + \frac{4}{3a^{2}_{0}}\right)
\right]^{3/2}},\nb\\
\rho_{c}(\tau) &=& \frac{6\epsilon^{0}_{c}\epsilon_{c}}{\kappa_{5}}
\left[\tau\left(\tau + \frac{4}{3a^{2}_{0}}\right)
\right]^{-1/2},  
%(w = -2/3),
\eqn
from which we can see that the universe starts to expand from the singular
point $\tau = 0$. In this case the acceleration of the expansion is always 
negative.

On the other hand,   it can be shown that
\bqn
\lb{4.37a}
T_{c}(\tau) &=& a^{3}_{0}\left[\tau\left(\tau + \frac{4}{3a^{2}_{0}}\right)
\right]^{3/2},\nb\\
Y_{c}(\tau) &=& \epsilon^{0}_{c} a^{3}_{0} \tau^{2}
\left(\tau + \frac{2}{a^{2}_{0}}\right).
\eqn
Then, we find that $T_{c}(\tau)  - \left|Y_{c}(\tau)\right|
\rightarrow + \infty$, as $\tau \rightarrow  \infty$. Hence, the 
trajectories of  the visible and invisible 3-branes in this case are 
also given by Fig. 12 in the ($T, Y$)-plane.

When $w \not= - \frac{2}{3}$, Eq.(\ref{4.34}) has the solution,
\bqn
\lb{4.38}
& & X^{1+m} F\left(\frac{1}{2}, \frac{1+ m}{2m}; \frac{1+3m}{2m}; 
- \left(\frac{X}{X_{0}}\right)^{2m}\right) \nb\\
& & \;\;\;\;\; = \frac{1+m}{\sqrt{m}}\tau,
\eqn
where $X_{1} \equiv X^{-2m}_{0}$. Then, it can be shown that
\bq
\lb{4.38a}
X(\tau) = \cases{0, & $\tau = 0$,\cr
\infty, & $ \tau \rightarrow \infty$,\cr}
\eq
and that
\bqn
\lb{4.38b}
\rho_{c}(\tau) &=& \epsilon^{0}_{c}\epsilon_{c}
\left(\frac{12X_{1}}{\kappa^{2}_{5} 
(1-|w|)(3|w|-1)}\right)^{1/2}\frac{1}{X},\nb\\ 
\ddot{a}_{c}(\tau) &=& - \frac{\beta X^{\frac{6|w| -5}{3(1-|w|)}}}
{3(1-|w|)(3|w|-1)}\left\{ X^{-\frac{2(3|w| - 1)}{3(1-|w|)}}
\right.\nb\\
& & - X_{1}\left(3|w| -2)\right\}.
\eqn
Therefore, when $-1< w < -2/3$, there exists a moment, $\tau_{f}$, after
which $\ddot{a}_{c}(\tau)$ becomes positive, that is, the universe turns to
its accelerating phase from a decelerating one, where $\tau_{f}$ is the root
of the equation,
\bq
\lb{4.38c}
X^{-\frac{2(3|w| - 1)}{3(1-|w|)}} - X_{1}\left(3|w| -2\right) = 0.
\eq

On the other hand, it can be shown that
\bqn
\lb{4.38ca}
Y_{c}(\tau) &=& \epsilon^{0}_{c} \frac{3\beta^{3}X_{1}^{1/2}}{2-3w}
                X^{\frac{2-3w}{3(1+w)}}\nb\\
& & \times F\left(\frac{1}{2}, \frac{3w-2}{2(1+3w)}; 
    \frac{9w}{2(1+3w)}; 
- \left(\frac{X}{X_{0}}\right)^{2m}\right),\nb\\
T_{c}(\tau) &=& \beta^{3}X^{\frac{1}{1+w}},
\eqn
from which we find that $T_{c}(0) = 0 = Y_{c}(0)$, and
\bqn
\lb{4.38cb}
T_{c}(\tau) &\sim&  X^{\frac{1}{1+w}} \rightarrow \infty,\nb\\
\epsilon^{0}_{c} Y_{c}(\tau) &\sim& X^{\frac{2}{3(1+w)}} 
\rightarrow \infty,
\eqn
as $X \rightarrow \infty$ (or $\tau \rightarrow \infty$). Following the 
arguments given above it can be shown that in this
case the trajectories of the two 3-branes are given by Fig. 12, too.

\subsubsection{$w = - 1$}

In this case, introducing the function $X(\tau)$ via the relation
\bq
\lb{4.17}
a_{c} = e^{X},
\eq
we find that Eq.(\ref{4.14}) can be written as
\bq
\lb{4.18}
\ddot{X} + \frac{8}{9}e^{-4X} = 0,
\eq
which has the first integral, 
\bq
\lb{4.19}
\dot{X} = \pm \frac{2}{3} \sqrt{e^{-4X} + X_{0}},
\eq
where $X_{0}$ is an integration constant.  When $X_{0} = 0$
it can be shown that the corresponding solution represents a
vacuum 3-brane ($\rho_{c} = 0$), as that given by Eq.(\ref{4.15})
with $a_{0} = \sqrt{3/4}$. When $X_{0} < 0$ there are no real 
solutions. When $X_{0} > 0$, Eq.(\ref{4.19}) has the solution,
\bq
\lb{4.20}
a_{c}(\tau) = a_{0} 
\sinh^{1/2}\left(\frac{4}{3a^{2}_{0}} \tau\right).
\eq
Then, we find that
\bqn
\lb{4.21}
\rho_{c}(\tau) &=& - p_{c} = \frac{4\epsilon^{0}_{c}\epsilon_{c}}
{\kappa_{5} a^{2}_{0}},\nb\\
\ddot{a}_{c}(\tau) &=& 
\frac{4}{9a^{3}_{0}\sinh^{3/2}\left(\frac{4}{3a^{2}_{0}} \tau\right)}
\nb\\
& & \times
\left\{\sinh^{2}\left(\frac{4}{3a^{2}_{0}} \tau\right) - 1\right\},
\eqn
which shows that at the beginning the universe is decelerating, but 
after the moment $\tau = \tau_{f}$ it starts to accelerate, 
where $\tau_{f}$ is given by 
\bq
\lb{4.22}
\tau_{f} = \frac{3a^{2}_{0}}{4}\ln\left(1 + \sqrt{2}\right).
\eq
It is interesting to note that in the present case we have 
$\rho_{c} = - p_{c} = Const.$

On the other hand, from Eqs.(\ref{4.2aa}), (\ref{4.11}) and (\ref{4.20})
we find that
\bqn
\lb{4.22aa}
Y_{c}(\tau) &=& 2\epsilon^{0}_{c}a_{0} 
 \int{\sinh^{3/2}\left(\frac{4}{3a^{2}_{0}}\tau\right)d\tau},\nb\\
T_{c}(\tau) &=&  a_{0}^{3}  
    \sinh^{3/2}\left(\frac{4}{3a^{2}_{0}}\tau\right).
\eqn
Clearly, in this case we have $T_{c}(0) = 0 = Y_{c}(0)$, and
$T_{c}(\tau) - \left|Y_{c}(\tau)\right| \rightarrow - \infty$, as 
$\tau \rightarrow \infty$. Then, the trajectory of the visible 
3-brane in the ($T, Y$)-plane is given by the solid curve $OA$
for $\epsilon^{0}_{c} = +1$  and   $OB$ for 
$\epsilon^{0}_{c} = -1$ in Fig. 13. Similarly, the  trajectory 
of the invisible 3-brane is given by the dotted curves $OA$
and $OB$, respectively, for $\epsilon^{0}_{c} = +1$  and   
$\epsilon^{0}_{c} = -1$. The bulk is  the region between the solid
and dotted curves.

%%%%%%%%%%%%%%%%%%%%%%%%%%%%%%%%%%%%%%%%%%%%%%%%%%%%%%%%%%%%%%%%%%%%%%%%%%%%%%%
\begin{figure}
\includegraphics[width=\columnwidth]{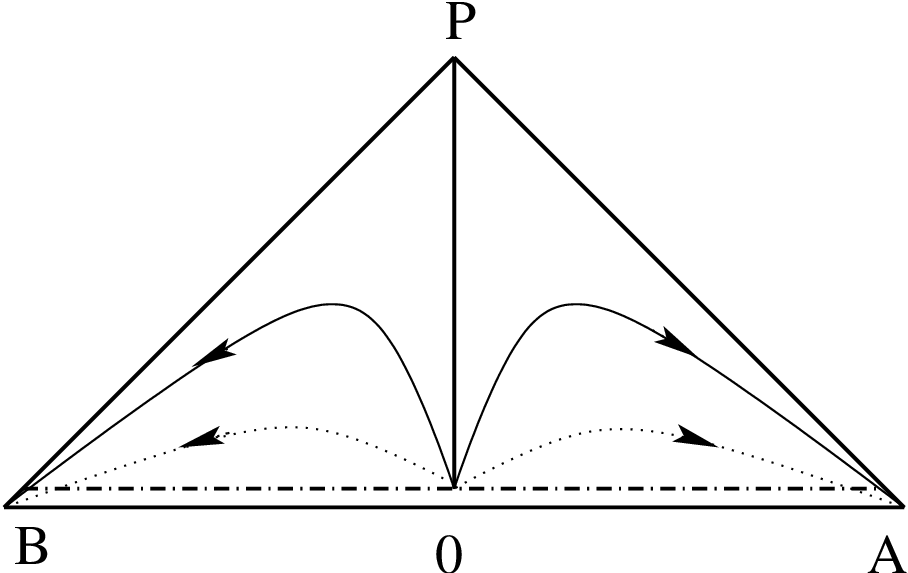}
\caption{The Penrose diagram of the solutions of Eq.(\ref{4.20}) 
in the $(T, Y)$-plane for $w = -1$.  The solid (dotted) curves $OA$  
and $OB$ are the trajectories of the visible (invisible) 3-branes, 
respectively, for $\epsilon^{0}_{c} = +1$ and $\epsilon^{0}_{c} = -1$. }  
\label{fig13} 
\end{figure} 
%%%%%%%%%%%%%%%%%%%%%%%%%%%%%%%%%%%%%%%%%%%%%%%%%%%%%%%%%%%%%%%%%%%%%%%%%%%%%%%

\subsubsection{ $ w < - 1$} 

In this case, choosing the
constant $\beta$ as that given by Eq.(\ref{4.26}), Eq.(\ref{4.25}) becomes 
\bq
\lb{4.39}
\ddot{X} - X^{\delta} = 0,
\eq
but now with
\bq
\lb{4.40}
\delta \equiv \frac{1 + 3|w|}{3(|w| - 1)}.
\eq
Eq.(\ref{4.39}) has the first integral,
\bq
\lb{4.41}
\dot{X} = \pm \sqrt{\frac{X^{2l} + X_{0}}{l}},
\eq
where 
\bq
\lb{4.42}
l \equiv \frac{3|w| -1}{3(|w| - 1)}.
\eq
To have an expanding universe, now we must choose the ``-"
sign in Eq.(\ref{4.41}). Then, Eq.(\ref{4.41}) has
 the solution,
\bq
\lb{4.43}
X \; F\left(\frac{1}{2}, \frac{1}{2l}; 1 + \frac{1}{2l}; 
- \left(\frac{X}{X_{0}}\right)^{2l}\right) = 
\left(\frac{X^{2l}_{0}}{l}\right)^{1/2}
\left(\tau_{s} - \tau\right),  
\eq
where $X_{1} \equiv X_{0}^{2l}$, and
\bq
\lb{4.44}
\tau_{s} \equiv \sqrt{\frac{l}{\pi}} X^{1-l}_{0}
\Gamma\left(1 + \frac{1}{2l}\right)
\Gamma\left(\frac{1}{2} - \frac{1}{2l}\right).
\eq
From  Eqs.(\ref{4.32b}) and
(\ref{4.43}) we find
\bq
\lb{4.45}
X(\tau) = \cases{0, & $\tau = \tau_{s}$,\cr
\infty, & $ \tau = 0$,\cr}
\eq
and 
\bq
\lb{4.46}
a_{c}(\tau) = \beta X^{-\frac{1}{3(|w|-1)}}
= \cases{\infty, & $\tau =  \tau_{s}$,\cr
0, & $ \tau = 0$.\cr}
\eq
On the other hand,   we also have
\bqn
\lb{4.47}
\rho_{c}(\tau) &=& \epsilon^{0}_{c}\epsilon_{c}
\left(\frac{12X_{1}}{\kappa^{2}_{5}
 (|w|-1)(3|w|-1)}\right)^{1/2}\frac{1}{X},\nb\\ 
\ddot{a}_{c}(\tau) &=&  \frac{\beta X^{-\frac{6|w| -5}{3(|w|-1)}}}
{3(|w|-1)(3|w|-1)}\left\{ X_{1} 
\right.\nb\\
& &\left. - X^{\frac{2(3|w| - 1)}{3(|w|-1)}}\right\},
\eqn
from which we find 
\bq
\lb{4.48}
\ddot{a}_{c}(\tau) = \cases{< 0, & $\tau < \tau_{f}$,\cr
>0, & $\tau > \tau_{f}$,\cr}
\eq
where $\tau_{f}$ is defined by,
\bq
\lb{4.49}
X(\tau_{f})  =  X^{\frac{3(|w|-1)}{2(3|w| - 1)}}_{1}.
\eq
Thus, in this case the universe  also has   an accelerating phase,
although at the beginning it is decelerating.
From Eq.(\ref{4.47}), on the other hand,  we find that
\bq
\lb{4.50}
\rho(\tau) = \cases{\infty, & $\tau = \tau_{s}$,\cr
0, & $\tau = 0$,\cr}
\eq
that is, in these models a big rip singularity develops
at $\tau_{s}$. In this case, it can also be shown that 
\bqn
\lb{4.51}
Y_{c} &=& -\frac{\epsilon^{0}_{c}\beta^{3}}{\pi^{1/2}}
X^{-\frac{1}{|w|-1}}_{0} \nb\\
& & \times
\Gamma\left(\frac{6|w|-5}{2(3|w|-1)}\right)
\Gamma\left(\frac{3|w|+2}{2(3|w|-1)}\right)\nb\\
& &  + \epsilon^{0}_{c}\beta^{3}X^{-\frac{1}{|w|-1}}\nb\\
& & \times  
F\left(\frac{1}{2}, \frac{3}{2(1-3|w|)};
\frac{6|w|-5}{2(3|w|-1)}; 
- \left(\frac{X}{X_{0}}\right)^{2l}\right),\nb\\
T_{c} &=& \beta^{3} X^{-\frac{1}{|w|-1}}.
\eqn
Thus, as $\tau \rightarrow 0$ (or $X \rightarrow \infty$), 
we have $T_{c}(\tau), \; Y_{c}(\tau) \sim 0$,  and
as $X \rightarrow 0$ (or $\tau \rightarrow \tau_{s}$) we have
\bqn
\lb{4.52}
& & \epsilon^{0}_{c}Y_{c}(\tau) \sim  \beta^{3} 
X^{-\frac{1}{|w|-1}} \rightarrow + \infty, \nb\\
& & T_{c}(\tau) \sim \beta^{3} X^{-\frac{1}{|w|-1}}
\rightarrow + \infty,  \nb\\
& & T_{c}(\tau) - \left|Y_{c}(\tau)\right| \sim 0.
\eqn 
Then, the trajectory of the visible 3-branes is given by the 
solid curve $OA$ for $\epsilon^{0}_{c} = +1$ and 
$OB$ for $\epsilon^{0}_{c} = -1$ in Fig. 14. Similarly, it can
be shown that the
trajectory of the invisible 3-branes is given by the 
dotted curve $OA$ for $\epsilon^{0}_{c} = +1$ and 
 $OB$ for $\epsilon^{0}_{c} = -1$. 
The spacetime of the five-dimensional bulk is the 
region between the solid and dotted curves. It is interesting
to note that the bulk is singular at the point $O$ ($T = 0$), 
but the spacetimes of the two 3-branes are not. On the other
hand, at the points $A$ and $B$ the spacetimes of the two 
3-branes have big rip singularities, but the bulk is regular 
there.

%%%%%%%%%%%%%%%%%%%%%%%%%%%%%%%%%%%%%%%%%%%%%%%%%%%%%%%%%%%%%%%%%%%%%%%%%%%%%%%
\begin{figure}
\includegraphics[width=\columnwidth]{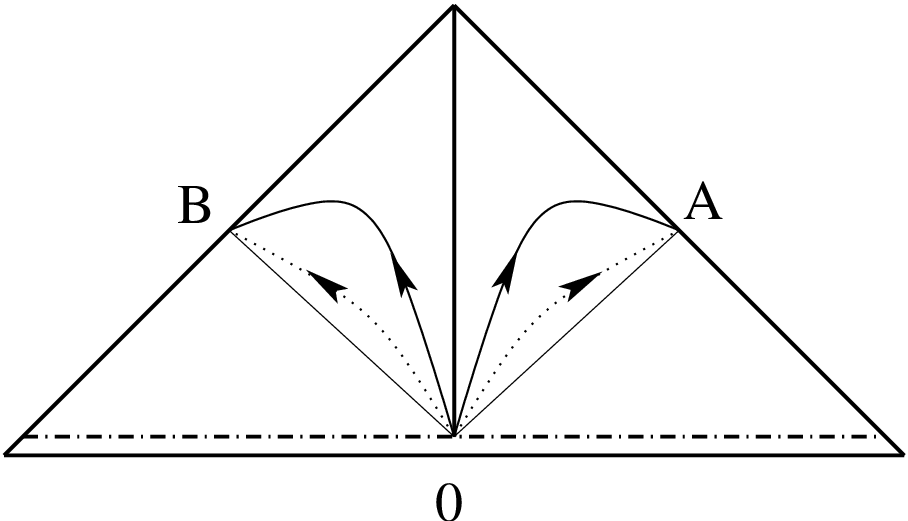}
\caption{The Penrose diagram of the solutions of Eq.(\ref{4.43}) 
in the $(T, Y)$-plane for $w < -1$.  The solid (dotted) curves $OA$  
and $OB$ are the trajectories of the visible (invisible) 3-branes, 
respectively, for $\epsilon^{0}_{c} = +1$ and $\epsilon^{0}_{c} = -1$. 
At the point $O$ the five-dimensional bulk is singular, but
the spacetimes of the two 3-branes are not, while at the points $A$ and
$B$ the spacetimes of the two 3-branes have big rip singularities, but
the bulk is regular.}  
\label{fig14} 
\end{figure} 
%%%%%%%%%%%%%%%%%%%%%%%%%%%%%%%%%%%%%%%%%%%%%%%%%%%%%%%%%%%%%%%%%%%%%%%%%%%%%%%

\section{Conclusions and Remarks}

In this paper, we have systematically studied cosmological models in the
 Randall-Sundrum setting of  two 3-branes \cite{RS1}. By assuming that   {\em 
the two orbifold 3-branes are spatially homogeneous, isotropic, and 
independent of time,} we have derived the most general form of metric
with orbifold symmetry, given by Eqs.(\ref{2.7b}) and (\ref{2.9}). Then, 
using distribution theory, we have developed the general formulas for 
two orbifold 3-branes. In particular, the field equations 
have been divided into three
sets, one holds in between the two 3-branes, given by Eqs.(\ref{2.7a})
and (\ref{2.18a}), one holds on each of the two 3-branes, given, respectively,
by Eqs.({\ref{2.15a}), ({\ref{2.18b}), ({\ref{2.15b}), and ({\ref{2.18c}).

Then, in Sec. III we have applied our general formulas  
to the case where the bulk is an $AdS_{5}$ space. Assuming that the cosmological
constant on the visible brane vanishes and that the equation of state of the
matter field takes the form, $p_{c} = w \rho_{c}$, with $w$ being a constant, 
we have been able to find the explicit expression for the expansion factor 
$a_{c}(\tau)$,  where $\tau$ denotes the proper time of the visible brane.
Although the curvature of the three spatial space of the brane is zero, it 
has been shown that for $w > -1$ the universe starts to expand from a big-bang 
like singularity at $\tau = 0$, until the moment $\tau = \pi/[6{\ell}(1+w)]$,
where $\ell \equiv (-\Lambda/6)^{1/2}$, and $\Lambda$ is the cosmological 
constant of the bulk. From that moment on the universe starts to contract until 
the moment  $\tau = \pi/[2{\ell}(1+w)]$, where the whole universe is crushed into 
a spacetime singularity. For $-1 < w < -2/3$ the universe experiences 
an accelerating period.  The global structure of the bulk and the embedding 
of the 3-branes in the
bulk have also been studied, and found that in some cases the geodesically 
complete spacetimes contain infinite number of 3-branes [cf. Fig. 8].

In Sec. IV, similar considerations have been carried out for the case where the 
bulk is vacuum. It has been shown that the vacuum solutions can be classified into
three different families, Class IA, Class IB, and Class II, given, respectively,
by Eqs.(\ref{4.1a}),  (\ref{4.1b}) and (\ref{4.2}). For Classes IA and IB solutions,
it has been found that the universe is always accelerating for $w < -2/3$, while
for Class II solutions with $w < -2/3$ the universe is first decelerating and 
then accelerating. The study of the global structure of the bulk as well as the
3-branes have shown that some solutions may represent the collision of two orbifold
3-branes [cf. Fig. 10].

It would be very interesting to look for solutions where the bulk is filled with 
other matter fields. In particular, look for solutions where acceleration of the 
universe is realized without dark energy, similar to the one-brane case \cite{reviews}.

Another interesting application of our general formulas is to cyclic universe scenario 
\cite{ST02} to study the collision of two orbifold 3-branes and the cosmological constant 
problem \cite{lambda}.

%%%%%%%%%%%%%%%%%%%%%%%%%%%%%%%%%%%
%\newpage
\section*{\bf Acknowledgments}

One of the author (AW) would like to thank the Center of Astrophysics, 
Zhejiang University of Technology, for hospitality, and Baylor University
for Summer 2006 sabbatical leave. RGC was supported by a grant from Chinese
Academy of Sciences, and grants from NSFC, China (No. 10325525 and
No.90403029).


\begin{thebibliography}{99}

\bibitem{ADD}  N. Arkani-Hamed, S. Dimopoulos and G. Dvali, Phys. Lett. {\bf
 B429}, 263 (1998); Phys. Rev. {\bf D59}, 086004 (1999); I. Antoniadis, N.
 Arkani-Hamed, S. Dimopoulos and G. Dvali, Phys. Lett., {\bf B436},Â  257 (1998).

\bibitem{RS1} L. Randall and  R. Sundrum, Phys. Rev. Lett. {\bf 83},
3370Â  (1999).

%\bibitem{RS2} L. Randall and  R. Sundrum, Phys. Rev. Lett. {\bf 83},
% 4690  (1999).

\bibitem{reviews} V.A. Rubakov, Phys. Usp. {\bf 44}, 871 (2001); 
%
S. F\"orste, Fortsch. Phys. {\bf 50},  221 (2002); 
%
C.P. Burgess, {\em et al}, JHEP, {\bf 0201}, 014 (2002); 
%
E. Papantonopoulos,  Lect. Notes Phys.  {\bf 592}, 458 (2002);
%
R. Maartens,  Living Reviews of Relativity {\bf 7} (2004); 
%  
P.~Brax, C.~van de Bruck and A.~C.~Davis,  
  Rept. Prog. Phys.  {\bf 67}, 2183 (2004) [arXiv:hep-th/0404011];
% 
U. G\"unther and A. Zhuk, ``{\em Phenomenology of Brane-World 
Cosmological Models,}" arXiv:gr-qc/0410130 (2004); 
%
P. Brax, C. van de Bruck, and A.C. Davis, ``{\em Brane World Cosmology}," 
   Rept. Prog. Phys. {\bf 67},  2183 (2004) [arXiv:hep-th/0404011]; 
%
V. Sahni, ``{\em Cosmological Surprises from Braneworld models of Dark Energy}," 
   arXiv:astro-ph/0502032 (2005); 
%
R. Durrer, ``{\em Braneworlds}," arXiv:hep-th/0507006 (2005); 
%
D. Langlois, ``{\em Is our Universe Brany}," arXiv:hep-th/0509231  (2005);
%
A. Lue, ``{\em Phenomenology of Dvali-Gabadadze-Porrati Cosmologies},"
   Phys. Rept. {\bf 423},  1 (2006) [arXiv:astro-ph/0510068];
%
D. Wands, ``{\em Brane-world cosmology}," arXiv:gr-qc/0601078 (2006); 
%
R. Maartens, ``{\em Dark Energy from Brane-world Gravity},"
   arXiv:astro-ph/0602415 (2006).



\bibitem{Hoyle04} C.D. Hoyle {\em et al.}, Phys. Rev. Lett. {\bf
86}, 1418 (2001); J. Chiaverini {\em et al.}, {\em ibid.}, {\bf
90}, 151101 (2003); J.C. Long {\em et al.}, Nature (London), {\bf
421}, 922 (2003); C.D. Hoyle {\em et al.}, Phys. Rev. {\bf D70},  042004
(2004) [arXiv:hep-ph/0405262].

\bibitem{DHR00} A. Pomarol, Phys. Lett. {\bf B486}, 153 (2000); 
H. Davoudiasl, J.L. Hewett, and T.G. Rizzo,  {\em ibid.}, {\bf B473}, 43 (2000);
Phys. Rev. Lett. {\bf 84},  2080  (2000); Phys. Rev. {\bf D63}, 075004 (2001);
E. Pree and M. Sher, {\em ibid.}, {\bf D73}, 095006 (2006) and references therein.

\bibitem{RC04} T. G. Rizzo, ``{\em Pedagogical Introduction to Extra Dimensions}, 
               arXiv:hep-ph/0409309 (2004); 
C. Cs\'aki, ``{\em TASI Lectures on Extra Dimensions and Branes}," 
   arXiv:hep-th/0404096 (2004).


\bibitem{GW99} Goldberger and Wise, Phys. Rev. Lett. {\bf 83},  4922  (1999);
Phys. Rev. {\bf D60}, 107505 (1999); C. Charmousis, R. Gregory, and V.A. Rubakov,
{\em ibid.}, {\bf   D62},  067505  (2000); C. Csaki, M. Graesser, L. Randall, 
and J. Terning, {\em ibid.}, {\bf   D62},  045015 (2000); 
B. Grinstein, D.R. Nolte, W. Skiba, {\em ibid.}, {\bf   D63},  105016 (2001); 
U. Gen and M. Sasaki, Prog. Theor. Phys. {\bf 105},  591  (2001);
G.D. Kribs, ``{\em  Physics of the radion in the Randall-Sundrum scenario},"
     eConf. C010630, P317  (2001) [arXiv:hep-ph/0110242]; 
J.E. Kim, B. Kyae, H.M. Lee, Phys. Rev. {\bf D66},  106004  (2002);
S. Bae, P. Ko, H.S. Lee, and J. Lee, ``{\em Radion Phenomenology in the Randall 
    Sundrum Scenario}," arXiv:hep-ph/0103187  (2001); 
A. Mazumdar and A. Pérez-Lorenzana, Phys. Lett. {\bf B508},  340 (2001);
Z. Chacko and P.J. Fox,  Phys. Rev. {\bf D64},   024015 (2001); 
P. Binétruy, C. Deffayet, and D. Langlois,  Nucl. Phys. {\bf B615},  219 (2001);
S. Kanno and J. Soda, Phys. Rev. {\bf D66}  083506 (2002); 
S. Nojiri, S.D. Odintsov, and A. Sugamoto,  Mod. Phys. Lett. {\bf A17},  1269 (2002);
P. Brax, C. van de Bruck, A.C. Davis, and C.S. Rhodes, Phys. Lett. {\bf B531},  135 
   (2002); 
S. Nasri, P.J. Silva, G.D. Starkman, and M. Trodden, Phys. Rev. {\bf D66},  045029
   (2002);
B. Grzadkowski, and J.F. Gunion, Phys. Rev.   {\bf D68},  055002 (2003);
S. Mukohyama and A. Coley, {\em ibid.}, {\bf D69},   064029 (2004);  
A. Mazumdar, R.N. Mohapatra, and A. Pérez-Lorenzana, JCAP, {\bf 0406},  004 (2004);
M. Toharia, Mod. Phys. Lett. {\bf A19},  37 (2004);
P.R. Ashcroft, C. van de Bruck, and A.-C. Davis,  Phys. Rev. {\bf D71},  023508 (2005);
C. Charmousis and U. Ellwanger,  JHEP {\bf 0402},  058 (2004); 
 G.L. Alberghi, D. Bombardelli, R. Casadio, and A. Tronconi, Phys. Rev. {\bf D72},  025005
    (2005); 
G.L. Alberghi and A. Tronconi, {\em ibid.},  {\bf D73}, 027702  (2006);
D. Konikowska, M. Olechowski, M.G. Schmidt,  {\em ibid.},  {\bf D73},  105018 (2006);
E.E. Boos, Y.S. Mikhailov, M.N. Smolyakov, and I.P. Volobuev, Mod. Phys. Lett. 
{\bf A21},  1431 (2006); A.S. Mikhailov, Y.S. Mikhailov, M.N. Smolyakov, and 
I.P. Volobuev, arXiv:hep-th/0602143  (2006); and references therein.
     
\bibitem{KI99} P.Kraus, JHEP,  {\bf 9912},  011 (1999); 
H.A. Chamblin, M.J. Perry and H. Reall, {\em ibid.}, {\bf 9909}, 014 (1999);
D. Ida, {\em ibid.},  {\bf 0009},  014 (2000); 
L. Anchordoqui, C. Nunez and K. Olsen,  {\em ibid.},  {\bf 0010},  050 (2000);
H.A. Chamblin and H. Reall, Nucl. Phys. {\bf B562},  133 (1999);
S. Mukohyama, T. Shiromizu and K.-I. Maeda,Phys. Rev. {\bf D62}, 024028 (2000);  
S.W. Hawking, T. Hertog and H.S. Reall, {\em ibid.}, {\bf D62}, 043501 (2000);
S. Gubser, {\em ibid.}, {\bf D63},  084017 (2001); Y. Wu, M.F.A. da Silva, 
N.O. Santos, and A. Wang, {\em ibid.}, {\bf D68}, 084012 (2003).


\bibitem{RS1CMs} C. Cs\'aki, M. Graesser, C. Kolda, and J. Terning, Phys. Lett.
     {\bf B462}, 34 (1999); 
C.Csaki, M.Graesser, C.Kolda and J.Terning, {\em ibid.}, {\bf B462}, 34 (1999);
T. Nihei, {\em ibid.}, {\bf B465}, 81 (1999);
J.Cline, C.Grojean and G.Servant, Phys. Rev. Lett. {\bf 83}, 4245 (1999);
N. Kaloper, Phys. Rev. {\bf D60}, 123506 (1999);
D.J.Chung and K.Freese, {\em ibid.}, {\bf D61}, 023511 (2000);
T. Shiromizu, K.-i. Maeda, and M. Sasaki,  {\em ibid.}, {\bf D62},  024012 (2000);
R.Maartens, {\em ibid.}, {\bf D62},  084023 (2000);
P. Bin\'etruy, C. Deffayet, U. Ellwanger, and D. Langlois,  Nucl. Phys. {\bf B477}, 
     285 (2000); 
P. Bin\'etruy, C. Deffayet,  and D. Langlois, {\em ibid.}, {\bf 565}, 269 (2000); 
    {\em ibid.}, {\bf B615}, 219 (2001); 
L. Mersini, Mod. Phys. Lett. {\bf A16}, 1583 (2001); 
H.-Y. Liu and P.S. Wesson, Astrophys. J. {\bf 562}, 1 (2001);  
A.N. Aliev, A.E. Gumrukcuoglu, Class. Quant. Grav. {\bf 21},  5081 (2004); 
C. de Rham and S. Webster, Phys. Rev. {\bf D71}, 124025 (2005); {\em ibid.},  
    {\bf D72}, 064013 (2005); 
C. de Rham, S. Fujii, T. Shiromizu, and H. Yoshino, {\em ibid.}, {\bf D72},  
    123522 (2005);  
T. Shiromizu, S. Fujii, C. de Rham, and H. Yoshino, {\em ibid.}, {\bf D73},  
    087301 (2006).

\bibitem{BCG00} P. Bowcock, C. Charmousis,  and R. Gregory, Class. Quantum Grav.
{\bf 17}, 4745 (2000).

 
\bibitem{Wolf} J.A. Wolf, {\em Spaces of Constant Curvature}, Fifth Edition (Publish 
or Perish, Inc., Wilmington, Delaware, U.S.A., 1984).



\bibitem{NoteA} A possible singular point of the transformation (\ref{2.6}) is the 
place where the two branes collide. 



\bibitem{Gib93} G.W. Gibbons, Nucl. Phys. {\bf B394}, 3 (1993).

\bibitem{CGS93} M. Cvetic, S. Griffies, and H.H. Soleng, Phys. Rev. {\bf D48},
2613 (1993); S. Griffies, ``{\em Field Theoretic and Spacetime Aspects of
Supersymmetric Walls and the Spacetime Aspects of Non-Supersymmetric Vacuum
Bubbles}," Ph.D. dissertation (University of Pennsylvania, 1993).

\bibitem{HE73} S.W. Hawking and G.F.R. Ellis, {\em The large scale structure
of space-time} (Cambridge University Press, Cambridge, 1973).

\bibitem{Taub51} A.H. Taub, Ann. Math. {\bf 53}, 472 (1951) [Reprinted in
Gen. Relativ. Grav. {\bf 36}, 2699 (2004)]; Phys. Rev. {\bf 103}, 454 (1956);
J. Ipser and P. Sikivie, Phys. Rev. {\bf D30}, 712 (1984).

\bibitem{ST02} J.~Khoury, B.~A.~Ovrut, P.~J.~Steinhardt, and N.~Turok, 
    Phys. Rev.  {\bf D64}, 123522 (2001);   
J.~Khoury, B.~A.~Ovrut, N.~Seiberg, P.~J.~Steinhardt, and N.~Turok,  {\em ibid.}, 
   {\bf D65},  086007 (2002);
P.J. Steinhardt and N. Turok,  {\em ibid.}, {\bf D65},  126003 (2002); 
%
 P.~J.~Steinhardt and N.~Turok, Science, {\bf 296}, 1436 (2002);
%
N.~Jones, H.~Stoica and S.~H.~H.~Tye, JHEP {\bf 0207},  051 (2002);
%
J. Khoury, P.J. Steinhardt, and N. Turok, Phys. Rev. Lett. {\bf 92},  031302 
(2004);  
%
J. Khoury, ``{\em A Briefing on the Ekpyrotic/Cyclic Universe}," astro-ph/0401579  
(2004);
%
N. Turok and P.J. Steinhardt, ``{\em Beyond Inflation: A Cyclic Universe Scenario}," 
 arXiv:hep-th/0403020 (2004);
%
A.~J.~Tolley, N.~Turok and P.~J.~Steinhardt, Phys. Rev. {\bf D69}, 106005  (2004);
%
Y.-S. Piao, {\em ibid.}, {\bf D70},  101302 (2004);
%
N.~Turok, M.~Perry and P.~J.~Steinhardt, {\em ibid.},   {\bf D70}, 106004  
(2004);  {\em ibid.}, {\bf D71},  029901 (Erratum)  (2005); 
%
Y.-I. Takamizu and K.-I. Maeda, {\em ibid.}, {\bf D70},  123514 (2004); 
     {\em ibid.},  {\bf D73},  103508 (2006);
%
G.~W.~Gibbons, H.~Lu and C.~N.~Pope, Phys. Rev. Lett. {\bf 94}, 131602  (2005); 
%
T.J. Battefeld, S. P. Patil, and R. H. Brandenberger, Phys. Rev. {\bf D73},  
086002 (2006);
%
A.J. Tolley, {\em ibid.}, {\bf D73}, 123522 (2006);
%
P.J. Steinhardt and N. Turok, Science {\bf 312},  1180 (2006); and references therein.


\bibitem{lambda} S. Weinberg, Rev. Mod. Phys. {\bf 61}, 1 (1989); S.M. Carroll,
``{\em The Cosmological Constant}," areXiv:astro-ph/0004075 (2000); T. Padmanabhan,
Phys. Rept. {\bf 380}, 235 (2003); S. Nobbenhuis, ``{\em Categorizing Different
Approaches to the Cosmological Constant Problem}," areXiv:gr-qc/0411093 (2004);
J. Polchiski, ``{\em The Cosmological Constant and the String Landscape},"
arXiv:hep-th/0603249 (2006); and references therein.
 


 


   

\end{thebibliography}
\end{document}